\documentclass[useAMS,usenatbib]{mn2e}
\usepackage{graphicx}

\def\la{\raise.5ex\hbox{$<$}\kern-.8em\lower 1mm\hbox{$\sim$}}
\def\ma{\raise.5ex\hbox{$>$}\kern-.8em\lower 1mm\hbox{$\sim$}}

\def\msol{M$_{\odot}$ }
\def\Lsol{L$_{\odot}$ }
\def\kms{$\rm km\, s^{-1}$}
\def\cm3{$\rm cm^{-3}$}
\def\Ts{$\rm T_{*}$~}
\def\Vs{$\rm V_{s}$~}
\def\n0{$\rm n_{0}$}
\def\B0{$\rm B_{0}$}
\def\ne{$\rm n_{e}$~}

\def\Te{$\rm T_{e}$}

\def\erg{$\rm erg\, cm^{-2}\, s^{-1}$}

\def\L12{L$_{12\mu m}$~}
\def\F12{F$_{12\mu m}$~}

\def\Hb{H${\beta}$~}
\def\Ha{H${\alpha}$~}
\def\Hg{H${\gamma}$~}

\def\Ly{Ly$\alpha$~}
\def\La{L$_{H\alpha}$~}
\def\Moy{M$_{\odot}$ yr$^{-1}$}

\def\ff{{\it ff}}

\def\RO3{R$_{[OIII]}$}

\title[SN and GRB host galaxy spectra]{
Physical conditions and element abundances  in  SN and GRB host galaxies
at different redshifts
}
\author[M. Contini]{ M. Contini
\\
School of Physics and Astronomy, Tel Aviv University, Tel Aviv
69978, Israel \\
}

\begin{document}


\pagerange{\pageref{firstpage}--\pageref{lastpage}} \pubyear{2009}

\maketitle

\label{firstpage}

\begin{abstract}

We compare the physical parameters and the relative abundances calculated  throughout supernova (SN) and
$\gamma$-ray burst (GRB) host galaxies
by the detailed modelling of the spectra.  The coupled effect of shocks and  radiation  from the starburst 
within the host galaxy is considered. We have found that
1) shock velocities  are lower in long period GRB (LGRB) than in SN host galaxies.
2)  O/H relative abundances in SN hosts are scattered within a  range 8.0 $<$12+log(O/H)$<$8.85
but they are  close to solar  in LGRB hosts.
 LGRB galaxies hosting Wolf-Rayet (WR) stars
have  He/H=0.13 in  a few objects.
3) The starburst temperatures within a few SN hosts   are  relatively high (\Ts$> 10^5$ K).  
\Ts in LGRB hosts are $\sim$ 3-8 10$^4$K.
4) The \Ha absolute flux calculated   from the emitting clouds of a few SN hosts
at  0.1$<$z$<$0.3 is sensibly higher than in the other galaxies.
 \Ha  increases sharply  with the ionization parameter $U$.
 The present analysis  suggests 
that the SN-host symbiosis is stronger than  for GRB in terms of activity.
The physical and chemical conditions in the GRB host galaxies are similar to those in SB galaxies within a large
z range.

\end{abstract}


\section{Introduction}

The explosion  of a degenerate carbon-oxygen white dwarf is at the origin of Type Ia Supernovae (SN).
SN Type Ib and Type Ic which  derive from the collapse of massive stars, 
(see Maoz et al 2014 for a review)
 were stripped of their H and He envelopes, respectively.
 SN Type IIn (e.g. Filippenko 1997) which are
 characterised by the presence of H, show a wide variety of photometric and spectroscopic
properties, while 
hydrogen deficient SN Type I  split into  Type Ia which show blended iron-group element emission lines, and
Type Ib and Type Ic which are characterised by several  lines of intermediate elements.
SN  explosions lead to collisions of the ejecta  with the host galaxy medium. 
A shock propagates in reverse through the circumstellar matter (CSM), while the  shock  accompanying
the ejecta  propagates outwards through the host galaxy clouds (Chevalier 1982). 

Observational and theoretical evidences indicate  that long duration $\gamma$-ray bursts (LGRB) 
 originate from the death of very massive stars
(e.g. Paczynski 1998), they are flashes of cosmic 
high energy ($\sim $1  ${\rm keV} - 10$ GeV) photons (Fishman \& Meegan 1995) and
 explode in star forming galaxies. 
 LGRB and their afterglows
 are  associated with broad lined SN Ic (e.g. Hjorth et al 2003, Stanek et al 2003).
They  provide  information about star-forming galaxies at high z
(e.g. Kr\"{u}hler et al 2015,  Blanchard et al 2015 and references therein).
 LGRB host galaxies are moderately star forming (SFR $\sim$ 1-10 \Moy), with stellar masses 
of $\sim$ 1-5  10$^9$ \msol 
and  high specific star formation rates (Fruchter et al 2006, Savaglio et al 2009). 
Moreover, they reveal 
 morphology, star forming rates and   burst energetics and location relatively to 
the galaxy centres (Castro-Tirado et al 2001). 
Sollerman et al (2005)  investigating low redshift  (z$<$ 0.2) GRB galaxy hosts  through  spectra
which include the HeI 5876 line, 
concluded that they are star-forming galaxies  (L$<$ \Lsol) with relatively low metallicity. 
Helium lines are   significant for GRB as well as  for SN Type Ic relatively to Wolf-Rayet (WR) stars,
which could be considered as LGRB progenitors, depending on  metallicity
(Han et al 2010). 

Short duration GRB (SGRB) last less than 2s, while LGRB have longer duration 
(Kouveliotou et al 1993). 
 Short bursts have significantly fainter afterglows. Their properties  are known from
their association with the host galaxies  (Fong et al. 2013). 
However, Fong et al could not find a trend linking GRB duration and host type.
The lack of an associated supernova and their link to a heterogeneous sample of host galaxies 
(e.g. Kann et al. 2011) is consistent with a compact binary merger origin
(Rosswog et al. 2003), such as neutron stars or black holes.
de Ugarte Postigo (2012) suggested that the merger is not associated with the galaxy star-forming region bulk,  
but it occurs in  dense regions.

In this paper we refer to  the narrow line-emission spectra from SN and GRB host galaxies.   In SN  hosts 
the  spectra  are emitted downstream of shock fronts   which   
 decelerate colliding with  high density clouds.
The line ratios account for the  physical conditions of the gaseous clouds, 
 for the photoionization flux from the starburst (SB) throughout the host and
 for the progenitor composition.
  The  spectra  seen   in GRB hosts 
 originate in star-forming regions, probing the conditions of the gas.
SN and GRB are generally  analysed by their light curves, whereas the host galaxy properties
are  investigated by the continuum spectral energy distribution (SED)
 in the different wavelength domains and by the line spectra.
 Spectroscopic data   provide   a full physical and chemical picture for local galaxies. At high redshifts
the  data are reduced to a few significant lines,  but the surveys contain hundreds of objects.
Therefore,  in order to obtain the O/H metallicity, 
O/H  are  calculated
by the "direct methods" (see e.g. Modjaz et al 2008).
The data are also  compared with diagnostic diagrams calculated for general cases.
On the other hand, the detailed  modelling of the spectra, which is currently used  to interpret  local galaxy spectra,
cannot be always adopted for   high z   objects, if the models
are  not constrained enough,  e.g. by the lack  of the auroral line [OIII]4363 (see Contini 2014a), 
one of the H lines such as \Hb, \Ha (\Hg is not always observed), He lines and others. 
This leads  in particular  to the  ambiguous identification of the gas
 photoionizing and heating sources (AGN, SB and/or shocks). 

In the following we   revisit the  emission  spectra from  the host galaxies of 
 super-luminous SN (SLSN)  presented by the Leloudas et al (2015) survey,
 of different SN types  by Sanders et al (2012),
of broad lined SN Type Ic   by Modjaz et al (2008),
of the LGRB hosts  from the surveys
 of Kr\"{u}hler et al,  Savaglio et al (2009),  Sollerman et al (2005), Castro-Tirado et al (2001), 
Graham \& Fruchter (2013), Levesque et al (2010), Vergani et al (2011), Piranomonte et al (2015), 
the LGRB line and continuum  spectra with WR features from the Han et al (2010) survey  and
the SGRB line spectra from de Ugarte Postigo et al (2014), Cucchiara et al (2013) and Soderberg et al (2006).
We have selected the spectra  suitable
to constrain  the models.
The code {\sc suma}\footnote{http://wise-obs.tau.ac.il/$\sim$marcel/suma/index.htm}
which simulates the physical conditions in a gaseous cloud under the coupled effect of
photoionization from a primary radiation source and shocks   was adopted to calculate the spectra. 
The line and continuum emission
from the gas are calculated consistently with dust-reprocessed radiation in a plane-parallel geometry.
The input parameters  refer to both the shock and the photoionization source.
We  obtain  the  physical characteristics of the different host galaxies  reproducing by the calculations  
the observed line ratios  of each object,    i.e.
  the photoionization source intensity flux, its  frequency distribution (AGN or SB)
and in the SB case  the effective temperature and ionization parameter. 
Moreover, we obtain the shock velocity, the preshock density,
 the geometrical thickness of the gaseous clouds and the element abundances throughout the host galaxy.

In the following, we  present new results  and compare  them  with those of previous investigations, 
in particular about the element abundances
at relatively high redshifts, because 
 metallicity is one of the fundamental parameters  which  affects the evolution of massive stars
(Piranomonte et al, Sollerman et al, etc).
In Sect. 2 we briefly  comment  about modelling approaches. In Sect. 3 we present the  modelling
of the galaxy hosts  and in Sect. 4  we discuss  the results regarding  the
SFR connection with the different physical parameters and element abundances. 
Concluding remarks  follow in Sect. 5.

\section{Modelling approach}

 The code calculates the emission line and continuum flux from a cloud of gas (and
dust), independently from the number of the observed lines.
The calculated line wavelengths range from far-UV to far-IR. We do not  report in the tables the calculated lines
which cannot be compared with the data.
We refer only to  spectra emitted  from the  SN and GRB host galaxies where  the different shock and 
photoionization events occur.
Detailed modelling by the codes e.g. CLOUDY, MAPPINGS, etc., that were assembled following 
the  observations of spectra rich in number of lines from different elements and in
different ionization stages, is  generally  disregarded for high z galaxies, because the
observed lines are few, while hundreds of galaxies are observed in each
survey.
Galaxies at high redshifts  often  originate from mergers and show a disturbed hydrodynamic structure.
Collisional phenomena are critical in the calculation of the spectra adopting 
models based on the coupled effect of photoionization and shocks. 
We suggest  that they are the closest approximation to
the  complex  structure of the emitting gas and they  are  therefore  suitable to SN and GRB host galaxies.

Different results derive from the   different interpretation methods.
Anderson et al (2016)  claim that emission line diagnostics are separated in two groups.
Empirical methods, using the [OIII] 5007/4363 ratio to estimate the emitting gas temperature,
calculate the O/H relative abundance directly from the oxygen line ratios to \Hb.
The second group compares the observed line ratios with those predicted by
photoionization/stellar population synthesis models.
Our approach is different.
A comparison between the results obtained by the "direct methods" and by the  detailed modelling  adopting SUMA
was presented by Contini (2014a) 
who   suggested  that the relative abundances obtained by
direct methods are lower limits, because  a constant temperature
is  adopted throughout all the galaxy.
 It is well known (Williams 1967) that the gas recombines following
the cooling rate by free-free, free-bound and line emission in the regions
farther from the radiation source and (Cox 1972)  following the cooling
rate downstream of  shock fronts. Therefore, regions of low temperature and  density in the galaxy
should be considered.
The line fluxes are calculated multiplying the fractional abundance of the
corresponding ions by the element abundance and integrating throughout the galaxy. The
fractional abundance of certain ions (e.g. O$^{++}$/O),  considering gas recombination, is lower than the
fractional abundance calculated  by an homogeneous temperature adapted to 
the [OIII]5007/4363 line ratio. So,  to reproduce  e.g. [OIII]/\Hb  a relatively high O/H
is adopted. 
 Moreover, all the line ratios in  each observed spectrum are reproduced by the same model.

Most of  the spectral  observations of  single high redshift  galaxies  cover the entire object.
 Interpreting the  average spectra by   models used to  calculate the  parameters in the different
regions  of a single local galaxy,
 we reach approximated or even  distorted  information.
For instance, the Galaxy is filled  with star clusters, colliding clouds of  dust and gas  
heated and ionized by the flux
from an active galactic nucleus (AGN),  from  SB, SN, etc. Turbulence leads to fragmented gas  clouds.  
The SN remnants (SNR) throughout the ISM
reveal   different physical conditions and element abundances, e.g. the higher than solar (see Table 1)
 He/H $\sim$0.4 in the Crab Nebula filaments  (Williams 1967, Contini et al. 1977),
the depletion of the heavy elements  in the Cygnus Loop (Contini \& Shaviv 1980), 
the different kinds
of H, He and N rich flocculi or O and S rich  knots  (Contini 1987) and other 
filaments (Fesen \& Milisavljevic 2015) in Cassiopeia A, the
dusty  clouds  showing  higher than solar N/H  in the Kepler SNR (Contini 2004), etc.
The  emitting clouds  surrounding  the explosion site are compressed, heated and ionized by the 
expanding shock 
which  blends with the ISM at large distances.
Shifting, for instance,  the Galaxy complex  to  a high redshift  and analysing the  average spectrum observed
at Earth (old position),  the  single region characteristic parameters   would result  
smoothed and averaged.

 The rich spectra observed from local galaxies allow to assemble a
multi-cloud model taking into consideration many different 
clouds within the galaxy (e.g. Rodriguez-Ardila et al 2005).
Actually, for high z objects, the observed spectra cover the entire galaxy
so, we select for each object the model  best fitting the line ratios of a prototype cloud which should prevail
throughout the galaxy.

\subsection{The calculation code}

By  SUMA, the
 calculations start at the shock front where the gas is compressed and thermalized adiabatically,
reaching the maximum temperature in the immediate post-shock region 
($T(K)\sim 1.5\times 10^5/(V_{\rm s}/100$ km s$^{-1}$$)^{2}$,
where \Vs is the shock velocity).
T decreases  downstream following the cooling rate.
This region  is cut into a maximum of 300 plane-parallel slabs
with different geometrical widths, which are
determined automatically, in order to  follow smoothly the temperature gradient.

The input parameters such as the  shock velocity \Vs, the atomic preshock density \n0,
the preshock magnetic field \B0, define the hydrodynamical field. They  are  used in the calculation
of the Rankine-Hugoniot equations  at the shock front and downstream. They  are combined in the
compression equation which is resolved  throughout each slab of gas in order to obtain the density 
profile downstream.
The input parameters  that represent the primary radiation field
for a   SB are  the effective temperature \Ts 
 and the ionization parameter $U$.
For an AGN, the primary radiation is the power-law radiation
flux  from the active center $F$  in number of photons cm$^{-2}$ s$^{-1}$ eV$^{-1}$ at the Lyman limit  and
spectral indices  $\alpha_{UV}$=-1.5 and $\alpha_X$=-0.7.
 The secondary diffuse radiation emitted from the slabs of gas heated by the shocks
is also considered.
Primary  and  secondary radiations are  calculated by radiation
transfer throughout the slabs downstream.
The dust-to-gas ratio ($d/g$) and the  abundances of He, C, N, O, Ne, Mg, Si, S, A, Fe, relative to H,
are also accounted for. They  affect the calculation of the cooling rate.
The input parameters (\Vs, \n0, \Ts, $U$, $D$, relative abundances, etc) which define a model, are the most
significant  ones in order to represent the  conditions of the gas
within  a specific object clouds. Hundreds of other parameters are considered in the
calculations  e.g. the recombination coefficients for each ion of each element,
the ionization cross sections etc.

Briefly, 

1)     an initial  input parameter set is adopted on the basis of the galaxy observations;

2)  the code calculates the density in the  slab of gas downstream  by the compression equation,

3)    the fractional abundances of the ions from each level for each element,

4)    line emission, free-free  and free - bound emission flux; 

5)    the temperature of the gas in the slab is recalculated by thermal balancing or the enthalpy equation; 

6)    the optical depth of the slab and the primary and secondary fluxes are calculated;

7)   the parameters found in slab i are  adopted as initial conditions  in slab i+1;

8)  integrating  on the line intensity increments calculated in each slab,  the
absolute flux of each line is obtained at the nebula (the same for the bremsstrahlung);

9)   the line ratios to a certain line (in the present case \Hb) are calculated.

\subsection{Method of analysis}

 In the modelling process we calculate  grids of models which  approximately reproduce the
observed line ratios. 
We start  adopting solar abundances as a first trial (Allen 1976, Table 1).
We  vary the input parameters until a satisfactory fit to 'all' the observed line ratios
is obtained consistently.
Each line has a different  strength which  translates in   different precision  by the fitting process.
A minimum number of significant lines (e.g [OIII] 5007+,  [OII]3727+, [OIII]4363, [NII], \Hb, \Ha) is 
necessary to constrain the model
but the  number of the observed lines  does not interfere with the modelling process. 
Each model calculates more than 200 line (and continuum) fluxes 
from far-UV to far-IR.
We deal with  line ratios to avoid  distance and morphological effects.
 We start referring to   line ratios of the same element (e.g. [OIII]/[OII], HeII/HeI) 
because they   depend  on  the physical conditions of the gas.
For  all models \B0=10$^{-4}$Gauss is adopted.
The models are initially constrained by
the observed FWHM of the line profiles (when available) which are roughly related  with  the  velocity field 
and can give a first hint to the shock velocity.
In a single galaxy, different emitting regions  can be directly recognized  from  complex FWHM.
This determines whether a pluri-cloud model should be considered.

Then,  the grids are  completed by comparing calculated with observed  line ratios to \Hb (e.g. [OIII]/\Hb),  
in order to obtain  the
relative abundances of the different elements. 
 If all the line ratios to H of a single element are higher (or lower) than the observed ones, the relative abundance
to H of that element is changed.
The abundances of strong coolants (e.g. He, C, O,)   affect the cooling rate
(by line emission) of the gas in the recombination zone. Consequently, by  increasing or reducing 
 one element abundance, all the calculated line ratios  change
and even if we have already reached a satisfactory fit  for  most of the line ratios, the whole modelling process 
is restarted until all the calculated line ratios are well tuned.
A perfect fit is not realistic because
the observed data have errors, both random and systematic. 
The uncertainty in the calculation
is due to  the atomic parameters (within 10 \%)  which are
often updated.
The strongest lines  (e.g. [OIII]5007)  are reproduced by $<$ 10\%, the weakest 
(e.g. [OIII]4363) by $\sim$ 50 \%. 
The calculation code and our modelling method  are described  by Contini (2014a, 2015 and references therein).

Different grids of models are calculated for different galaxy  types.
Dealing with spectra of  high z objects, which contain  mainly  oxygen lines,
a first  hint about the  model is  given by   [OIII]/[OII],
 which  does not depend on O/H, and from
 [OIII]/\Hb and [OII]/\Hb which depend on  both the physical conditions and O/H.
The ionization parameter $U$ which represents the ratio between the  number of photons from the SB in the host and the
number of  electrons in the cloud of gas reached by the radiation flux, links the  photoionization  source with the
 galaxy ISM properties.  Model  degeneracy  between \Ts and $U$
 is only apparent. For instance,
 a higher \Ts increases both the [OIII]/\Hb and [OIII]/\Hb line ratios but a higher $U$ not always yields a higher
[OII]/\Hb. 
  Line ratios to \Hb in the optical range, which are strong enough to be observed,
 are shown as  function of $U$ (top diagram) and \Ts (bottom diagram) in Fig. 1.
The chosen ranges are 10$^{-3}$ $\leq$$U$$\leq$1.  and 10$^4$$\leq$ \Ts $\leq$ 5.10$^5$ K
(Contini 2015 and references therein).
Such diagrams  suggest how to reproduce the observed spectra.

\begin{figure}
\includegraphics[width=8.8cm]{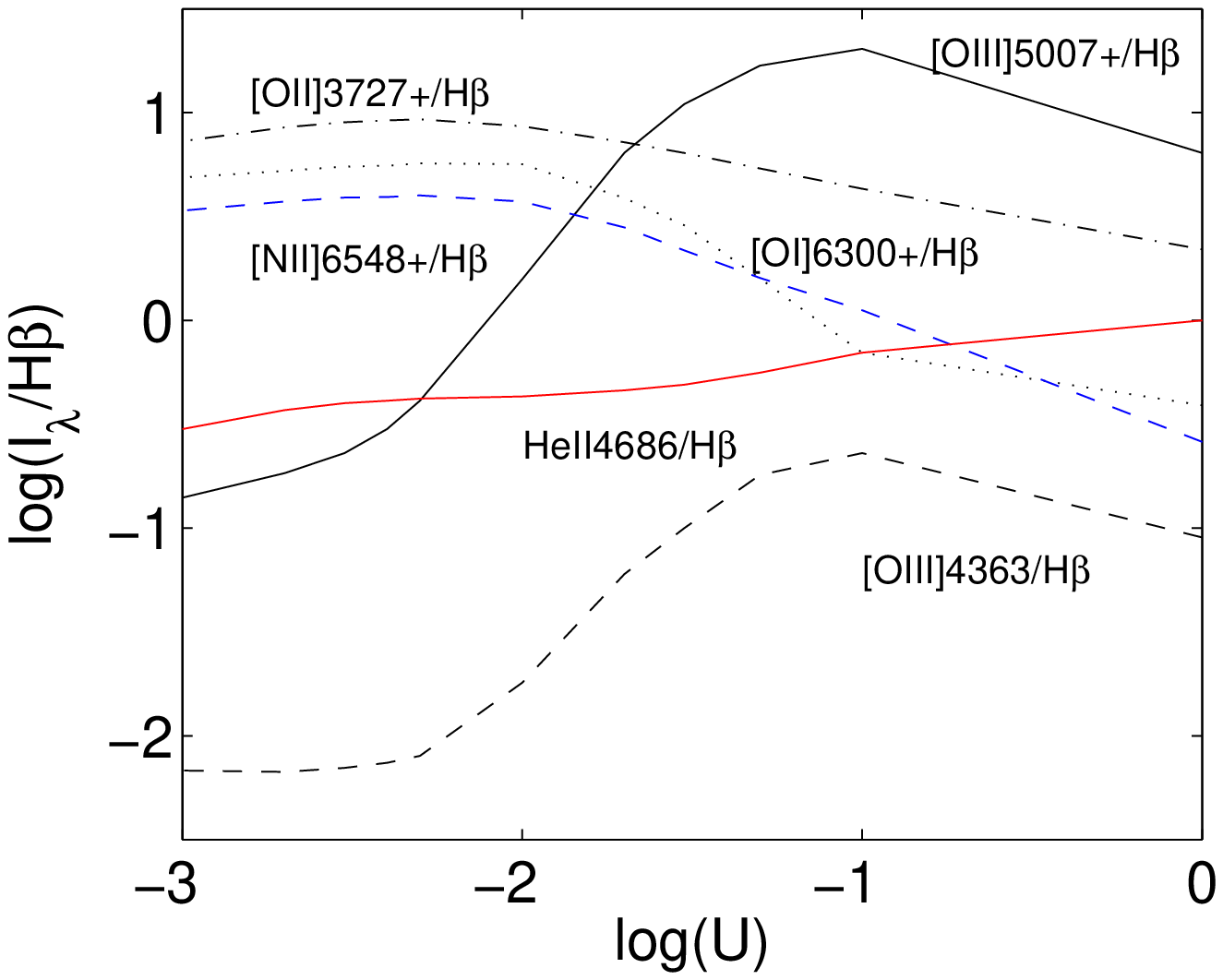}
\includegraphics[width=8.8cm]{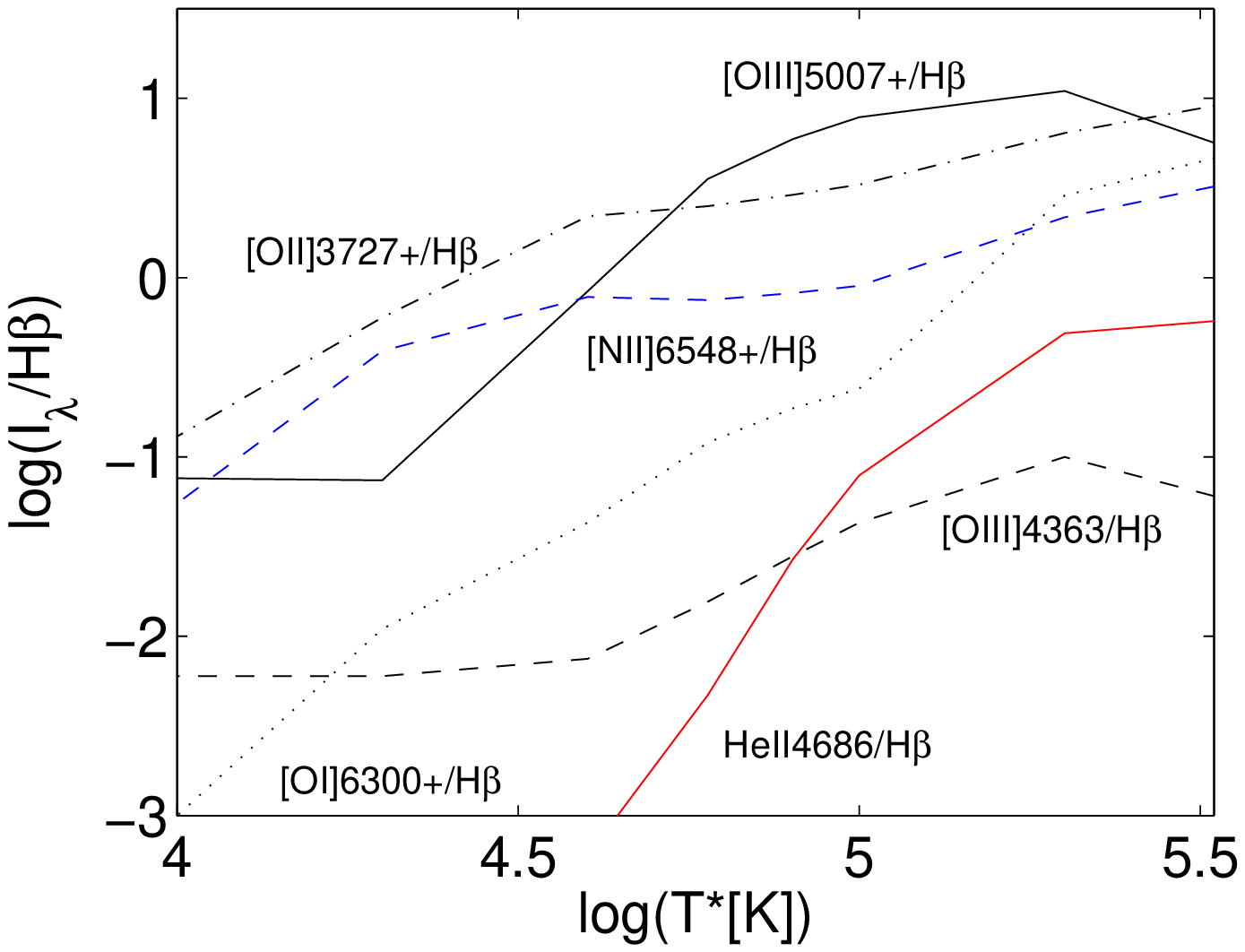}
\caption{
Top diagram. The line ratios to \Hb calculated as function of $U$  (\Ts=2 10$^5$ K).
Bottom diagram. The line ratios to \Hb calculated as function of \Ts  ($U$=0.03).
}
\end{figure}

\begin{table}
\caption{The solar element abundances}
\begin{tabular}{ccccccc} \hline  \hline
\ element & Allen & Anders \& Grevesse & Asplund et al.  \\
\         & (1976)       & (1989)      & (2009)      \\ \hline
\    H    &  12          &    12       & 12           \\
\   He    &  11          &  11         & 11          \\
\  C      &  8.52        &  8.56       &  8.43       \\
\  N      & 7.96         &  8.05       &  7.83       \\
\  O      & 8.82         &  8.93       &  8.69  \\
\   Ne    & 8.            & 8.09       & 7.93 \\
\ Mg      & 7.4          &  7.58       &  7.6 \\
\ Si      & 7.52         &  7.55       & 7.51   \\
\ S       & 7.2          &  7.21       & 7.12   \\
\ Cl      &  5.6         &   5.5       & 5.5   \\
\ Ar      & 6.52         &   6.56      & 6.4   \\
\ Fe      & 7.5          &   7.67      & 7.5  \\ \hline\\
\end{tabular}
\end{table}

\subsection{The density,  [SII] and  [NII] lines}

The gas density n is a crucial parameter in models  accounting for the shocks. 
In each cloud,  n reaches its upper limit  downstream and  remains nearly constant, 
while the electron density \ne decreases following recombination. 
Recall that n is linked with \n0 by 
 compression downstream (n/\n0) which ranges between $<$ 10 and $>$ 100, depending on \Vs and \B0.  
A high density, increasing the cooling rate,  speeds up the recombination process
of the gas,  enhancing the low ionization level lines. 
Each line comes from a region  of gas  at a different electron density  and temperature,  
depending on the ionization level and the  atomic parameters characteristic of the ion. 

 The density n is  roughly revealed by the
[SII]6716/6731 doublet ratio. 
The  [SII] lines
are also characterised by a relatively low critical density for collisional deexcitation.
In some cases   the  [SII]6716/6731 line ratio  varies from  $>$1 to  $<$1 
throughout a relatively small  region downstream. 
In fact   the [SII] line ratios depend on both the electron temperature 
and  density  of the emitting gas (Osterbrock 1974, fig. 5.3) which,  
in models accounting for the shocks, 
 are far from  constant throughout  the clouds.
So, even  sophisticated calculations which  reproduce approximately the highly inhomogeneous
conditions of the gas,  lead to some discrepancies between the calculated and observed line ratios.
So [SII]/\Hb ratios do not reproduce the observations 
with the required precision.
Unfortunately, there are no data for S lines from higher  ionization level, which could indicate whether
the choice of the  model is misleading or different relative abundances should be adopted.
Recall that sulphur can be easily depleted from the gaseous phase and trapped into dust grains and molecules.
We  consider that the [SII]/\Hb line ratios are indicative of the S/H relative abundance.

The same is valid for the [NII] lines.
The observations do not contain any other  strong line of N, which could confirm  whether the gap between 
the calculated and the observed [NII]/\Hb line ratios can be resolved by changing  the N/H relative abundance or
the other input  physical parameters.
Notice that [NII] lines come from the same region of the gas  that  emits the [OII] lines
due to charge exchange reaction between O$^+$,  N$^+$ and H$^+$.
[OII]/\Hb line ratios depend on the physical parameters and are constrained by [OIII]/\Hb,
so the [NII]/\Hb line ratios can  easily show abnormal N/H relative abundances.
Nitrogen is less abundant  than other elements e.g. oxygen, neon, etc., therefore it
 is not a strong coolant, namely, does not affect strongly the cooling rate of the gas downstream.
Changing N/H,  the [NII]/\Hb line ratios  accordingly change, but   the other line ratios are not
modified.

\subsection{O/H and the [OIII] 4363 line}

 O/H relative abundances calculated by  detailed modelling are generally higher than those
calculated by the "direct methods".
In fact, oxygen is a strong coolant which shapes 
the temperature, density profiles  and  the relative  distribution of the different ions throughout the host clouds.
The  contribution from   slabs of gas  to a line
 can be low in regions where the gas  conditions are less adapted to a high ion fractional abundance.
A  high element abundance  compensates the gap between the calculated and observed line intensity.
By the "direct methods" the regions of gas with temperatures and densities less adapted to a line 
are not considered and the element abundances are calculated directly considering the most
appropriated conditions for  the ion.

 A higher O/H  not always yields  higher  [OIII]/\Hb, 
[OII]/\Hb and [OI]/\Hb  by the same factor, sometimes it even leads  to  opposite trends, depending on 
the set of all the input parameters.
The [OIII]4363 line depends on the gas temperature and  density, constraining the model.
The final choice of a model is dictated by  [OIII]4363/\Hb (when observed,  see Contini 2016)
that is calculated consistently with [OIII]5007+/\Hb and [OII]/\Hb.
However, this line is generally  weak. 
Its role in modelling the spectra was discussed by Contini (2014a).

\section{Comparison of observed  with calculated spectra}

We have gathered the spectra emitted from  SN and GRB host galaxies. 
Modelling results of  selected objects  are  shown in the following.

\subsection{Leloudas et al (2015) SN galaxy host survey}

Leloudas et al (2015) present the spectra from the host galaxies of super luminous SN (SLSN). 
They are very bright explosions which mostly occur in faint dwarf galaxies.
The subtypes SLSNI and SLSNR are H poor and they are distinguished by  their light curve evolution.
SLSNII are H rich.  
Leloudas et al (2015) refer   to Extreme Emission Line Galaxies (EELG) observations.
For spectroscopy they used FORS2 and X shooter on the VLT, OSIRIS on the GTC and IMACS on Magellan.
Narrow lines
 originate from the  host galaxy, downstream of the shock produced by collision of the
 fast moving ejecta  with the CSM (see  Leloudas et al 2015
 and references therein).

We report in Table 2 the  data presented by Leloudas et al (2015) in their table B1. For each galaxy 
the  observed line ratios are followed in the  two next rows by the best fitting  results
calculated by shock dominated (SD) models (ml1-ml26) and radiation dominated (RD) models 
(mll1-mll26) which account  for  the photoionization flux from the SB + shocks, respectively.
In Table 2, columns 8-15, the  model input parameters  are reported.
The SD models (ml1-ml26),  which best reproduce  the [OII]/\Hb and [OIII]5007+/\Hb line ratios 
(the + indicates that the doublet is summed)
show relatively high shock velocities (except for SN 1999as and SN 1999bd). 
SD models are presented for all the objects even if a broad component was observed only in the 
 \Ha and [OIII] line profiles of the PTF11dsf galaxy. 
This broad component can be attributed to strong outflows  of SN winds (Leloudas et al) or  from SB star outbursts
(Contini 2015).
The relatively low velocities of RD models refer to the encounter  of the SN outflowing shock with the
ISM clouds. 
The shock velocities and pre-shock densities calculated by RD models are in  average similar to 
those in  the  proximity of 
SB clouds (Contini 2015 and references therein).
The results  of  absolute \Hb fluxes (in Table 2, col. 11) calculated for SD models
show that the intensity of the line fluxes  are lower  than those calculated by   coupled
photoionization and shock  by  factors between 10 and 1000. Therefore  the contribution
of high velocity clouds can be hardly distinguished from the underlying noise  of the spectra.
SD models    overpredict 
the observed [OIII] 4363/\Hb ratios by factors $>$3. Therefore, RD models are more appropriated.
Table 2   results  confirm that SN generally occur in SB galaxies 
and  show that some  star  temperatures are  approaching those characteristic of   outburst (see Contini 2014b).
The temperatures of a few SLSNI and SLSNR host SB are even higher than those found in galaxies 
with  activity at relatively high z.
We associate the “activity” in galaxies ( van Dokkum et al. 2005) to an AGN, a SB, and to shocks on 
the basis of a strong wind throughout the galaxy, while we refer to “quiescent” galaxies 
(Kriek et al. 2009) because  emitting weak lines. 
Those imply a low luminosity AGN and a low star formation rate (SFR = 1 - 3 \msol yr$^{-1}$) compared with models 
of stellar population synthesis SEDs. Contini (2014b)  claims that activity is connected with a high 
SB temperature (\Ts $>$ 10$^5$K). 
The O/H relative abundances are scattered between O/H = 10$^{-4}$ and 7.1 10$^{-4}$ (we refer to O/H solar
= 6.6 10$^{-4}$ by Allen 1976 in
Table 1).
 N/H  are generally lower than solar (10$^{-4}$) by  factors of 5-20.
The geometrical thickness of the emitting clouds  drops from $D$ $\leq$ 30 pc
for galaxies at z$\geq $0.3 to $\leq$0.3 pc at lower z. 
In general the shocks accompanying turbulence 
in the SB lead to fragmentation of  matter and to different
cloud thickness within  a large range.

\subsection{Sanders et al (2012) SN host galaxy survey}

Sanders et al presented the spectroscopic observations of a relatively large sample of different SN type
 host galaxies at 0.01$<$z$<$0.15. The authors were particularly interested  in the relative abundances of 
significant elements. The results of their investigation show rather low metallicities.
They were obtained by  different methods which are well described in their text.
These methods which  refer mainly to the oxygen lines cannot provide a consistent representation
of the physical properties of the emitting gas such as to explain 
all the line ratios reported by  each galaxy spectrum.
Therefore, we have revisited their spectra by  detailed modelling (msnd1-msnd24) in order  to compare the results and 
in particular, to provide a  more detailed  representation of the physical conditions throughout  
the emitting gaseous clouds.  The results of modelling  are presented in Table 3.
We constrain the models by the weakness of the [OIII]4363 line.
These spectra are well reproduced by clouds of gas propagating outwards from the SB (the photoionizing source), 
while the spectra
corresponding to relatively high [OIII]4363/\Hb ($\geq$ 0.1) are  reproduced by clouds propagating inward
towards the SB. In the latter case the radiation flux reaches the shock front of the cloud, in the former
case it reaches the edge opposite to the shock front.

To better understand this result we present in Fig. 2 the profile of the electron temperature, electron density
and fractional abundances of significant ions throughout the clouds for  models msnd15 and msnd15-0
corresponding to outflow and to inflow, respectively.
In the outflow case (top diagrams) the emitting cloud  is  divided into two halves represented by the left 
and right diagrams.  The left diagrams show
the region close to the shock front and the distance from the shock front on the X-axis scale is logarithmic.
The right diagrams show the conditions
downstream far from the shock front, close to the edge reached by the photoionization flux which is 
opposite to the shock front.  The distance from the
illuminated edge is given by a reverse logarithmic X-axis scale.
The inflow case is  represented by the bottom diagram. 
Fig. 2 diagrams   reveal why,
compared with Sanders et al results, the metallicities calculated by detailed modelling (Table 3) are generally 
higher than those deduced by different methods.
In the outflow case, the O$^{++}$/O fractional abundance is very low throughout more than half of the  geometrical
width of the cloud,
therefore to  obtain relative high [OIII]4363/\Hb a  relatively high O/H  is adopted (Sect. 2.4).
In the inflow case,  O$^{++}$/O is nearly constant and corresponds to T$\sim$ 10$^4$K, so a relatively low O/H= 4 10$^{-4}$
nicely fits the observed line ratio.
Outflow is more suitable to the clouds in SB environments. 
Inflow in few cases can be justified by the turbulent regime created by shocks throughout
SB regions.

\subsection{Modjaz et al (2008) broad-lined TypeIc SN host galaxies}

 The element abundances in the host  galaxies are  a basic issue  to understand
the characteristics of local and high redshift SN and GRB phenomena.
Modjaz et al (2008) presented   observations of 12 hosts of broad lined Type Ic
at z$<$0.14  with no observed GRB and derived host galaxy central metallicities (O/H)
and metallicities at the SN positions by strong-line diagnostics.
Comparing their results with those of five nearby SN-GRB hosts,
they found that broad lined SN Ic without GRBs show higher metallicity environments.

The new spectra presented by Modjaz et al 
are rich  of lines which permit the detailed modelling of the line ratios.
We have revisited  Modjaz et al spectra.
The host-galaxy observed emission  line-fluxes are given in their tables  4 and 5
measured in the galaxy centre and in the SN positions different from centre, respectively.
We  present in Tables 4 and 5 the modelling  (models mM1-mM8) of line ratios to \Hb
(corrected for extinction,  Osterbrock 1974).
The results reproduce the strong line data within 10\%. 
The observed sulphur doublet 6717/6731 is $>$1 
 which would correspond to a very low density (\ne$\leq$10 \cm3, adopting \Te=10$^4$K, Osterbrock 1974, fig. 5.3).
 Such densities are  low compared to those found in SN and GRB host galaxy emitting clouds.
We refer to preshock densities of $\sim$ 100 \cm3, which lead to the satisfactory fit of all the other line ratios.
This ambiguity is discussed in Sect. 2.3 and it is probably due to the  continuum subtraction.

The 12+(O/H) relative abundances which result from our modelling  range between  8.79 and 8.66 in the central 
galaxy regions
and  between  8.82 and 8.75  at the SN positions different from the galaxy centre.  Modjaz et al
 results range between a maximum of 9.15 and a minimum of 8.24 with different methods and for different galaxies
in the centre and between 9.08 and 8.38 in the SN position out of centre.

\subsection{Kr\"{u}hler et al (2015)  LGRB  host  galaxy survey}

 Kr\"{u}hler et al  investigate the physical conditions of LGRB host galaxies by recent observations
of  90 objects. 
 Long period GRB are connected with core-collapse of progenitors of SN of Type 1c.
 Kr\"{u}hler et al claim that a GRB explosion represents a very rare endpoint of stellar evolution.
They present VLT/X-shooter  emission lines of GRB-selected galaxies at 0.1$<$ z$<$ 3.6 and analyse the correlation
between the host physical properties and element abundances.
We revisited the spectra  
by the detailed modelling of the line ratios, using models which account consistently for photoionization 
from an SB + shocks. The spectra lacking the data  of some significant lines were excluded
because unable to constrain the models.  
In Table 6 we present the  results of modelling.
The line ratios have been reddening corrected.
The observed FWHM are calculated by  $\sigma$ =$\sqrt{FWHM^2-\Delta V^2}$/2$\sqrt{2ln2}$, where
$\Delta$V$\sim$ 35 \kms,   as given by Kr\"{u}hler et al. They are used to constrain  \Vs.
In Table 6, for each observed spectrum, the corrected line ratios are followed in the next row by model results
(mk1-mk52).
The  090201 spectrum, which shows particularly high gas velocities (400 \kms), could be reproduced by three 
different models, SD, SB and AGN dominated. The \Hb flux calculated by the AGN model is definitively higher
than those calculated by the other models,  suggesting that an AGN could be present in the host galaxy.
Compared with the models for SLSN hosts, long GRB hosts in the same z range 
show in general very low ionization parameters $U$,
indicating  relatively high  dilution of the radiation flux due to a large distance R of the emitting gas from
the photoionization source. The flux $F$ from the SB stars and the ionization parameter $U$
are combined  by $F$(r/R)$^2$=$U$nc, where r is the radius of the star, R the distance  to the  emitting nebula,
n the density of the gas and c the speed of light.
Alternatively, the flux is obstructed on its way to
the clouds. At higher z, the results are different (see Sect. 4.1).

\begin{table*}
\centering
\caption{Modelling  Leloudas et al (2015) SLSN  host spectra. (Line ratios to \Hb=1)}
\tiny{
\begin{tabular}{lccccccccccccccc} \hline  \hline
\  &z         & [OII]  & [OIII] & [OIII]  & \Ha & [NII] & \Vs  & \n0  & $D$          & \Hb & O/H       & N/H & \Ts & $U$       \\ 
\  &          & 3727+  & 4363   & 5007+   &6563 & 6584  & \kms & \cm3 & 10$^{18}$ cm & flux$^1$ &10$^{-4}$ & 10$^{-4}$ &10$^4$K & -  \\ \hline
\ SN1999as R&0.127& 9.14 & -     & 4.5     & 5.5  & 1.5   & -   &  -   &    -         &    -   &    -   &    -  &-&-   \\
\ ml1      && 8.8  & 0.37     & 4.67    & 4.5  & 2    & 200 & 50   &   1.43       & 0.00014 & 2.2   & 0.6 &-&-     \\    
\ mll1     && 8.4  & 0.05     & 4.5     & 3.   & 1.8  & 350 & 110  & 1.6          & 1.67    & 7.3   & 0.5 & 30 & 0.2 \\
\ SN as loc&0.127& $<$2.54&-   & 5.84    & 3.68 & $<$0.38& -  &  -   &  -           &   -     &  -    &  -  &  - &  -  \\
\ ml2     && 2.    & 0.34     & 6.1     & 3.2  & 0.2  & 760 & 110  & 0.87         & 0.0016 &  2     & 0.05&-&- \\
\ mll2    && 1.98  & 0.017    & 6.0     & 2.94 & 0.42 & 250 & 100  & 2.           & 0.084  &  5.    & 0.3 & 5 & 0.1   \\
\ SN1999bd II&0.151 & 3.4  & -   & 2.      & 4.64 & 1.44  &  -  &  -   &   -          &   -     &   -   &     -   &-&-  \\
\ ml3      && 3.3  & 0.16     & 2.      & 4.1  & 1.7  & 250 & 40   &  4.4         & 0.0002  & 1.2   & 0.6 &-&-    \\
\ mll3     && 3.2  & 0.005    & 1.98    & 3.   & 0.64 & 210 & 90   & 100      & 0.1     & 6.6   & 0.3   & 3.8  & 0.1  \\    
\ SN2006oz I&0.396& 3.7   &  -  & 5.      & 3.47  &$<$1.3& -   &   -  &    -         &  -     &   -    &  -&-&-   \\
\ ml4      && 3.9   &  0.2     & 4.6     & 3.   &0.22  & 740 & 230  & 0.4          & 0.012  & 5.1    & 0.04&-&- \\
\ mll4    && 3.5   & 0.012    & 4.87    & 3.   & 0.68 &220  & 90   & 100          & 0.1    & 6.4    & 0.3 & 5 & 0.1 \\
\ SN2006tf II&0.074&4.     & -   & 3.68    & 4.37 & 0.4   &   - &  -   &    -         &   -    &   -    &   - &-&-  \\
\ ml5      && 4.2   & 0.2     & 3.43    & 3.   & 0.4   &  600& 220  & 0.24         & 0.01   & 6.6    & 0.08&-&-  \\
\ mll5     && 4.4   & 0.053   & 3.54    & 3.9  & 0.8  & 340 & 120  & 1.6          & 1.27   & 5.     & 0.2 & 4 & 0.6\\
\ SNLS06D4eu I&1.588&2.33 & -   & 5.0     &2.32  & 0.24  &  -  &   -  &   -          &   -    &   -    &   - &-&-  \\
\ ml6       && 2.3  & 0.36    & 5.23    &3.46  & 0.28 &540  &190   & 0.243        & 0.002  &  2.    & 0.05&-&- \\
\ mll6      && 2.2  & 0.074   & 5.3     & 3.3  & 0.46 & 350& 120  & 25.        & 24.3    &  0.8   & 0.06  & 10. & 0.1  \\
\ SN2007bi R&0.128& 3.2  & -   & 4.17    &3.27  & 0.2   & -   &   -  &   -          &   -    &   -    &   -      \\
\ ml7       && 3.3  & 0.23    & 4.0     &3.1   & 0.39 & 650 & 75   & 1            & 0.02   & 3.6    & 0.1  &-&-    \\
\ mll7     && 3.2  & 0.026    & 4.17   & 2.94  & 0.4  & 240 & 130  & 0.4         & 0.026   & 3.    & 0.1   &7.3&0.02 \\
\ SN2008am II&0.233& 3.12 & -   & 3.      & 4.   & 0.56  &  -  &  -   &    -         &   -    &  -     &   -   &-&-    \\
\ ml8      && 3.   & 0.22     & 3.      & 3.2  & 0.76 &  500& 200  &    0.22      & 0.0045 &  3.9   &  0.2    &-&-   \\
\ mll8     && 3.   & 0.019    & 3.1     & 3.   & 0.75 & 240 & 130  & 0.4          & 0.026  &  3.    & 0.2   & 6.3& 0.02\\
\ PTF09cnd I&0.258& 4.12 & -   & 4.53    & 3.44 & 0.28  &   - &  -   &    -         &    -   &   -    &   -     &-&-   \\
\ ml10     && 3.9  & 0.2      & 4.6     & 3.   & 0.34  & 740 & 230  &   0.4        & 0.012  & 5.1    & 0.06  &-&-  \\
\ mll10    && 4.5  & 0.024    & 4.53    & 3.   & 0.65  & 240 & 130  &  0.4         & 0.026  & 5.9    & 0.2    & 6.3 & 0.02 \\
\ 2010gx I&0.230& 1.19   &  0.21   & 6.56    & 3.7 &  0.057& -    &   -  &  -           &   -  &    -     &     - &-&-     \\
\ ml11  && 1.04   & 0.45      & 6.87    & 3.  & 0.24  & 900  & 570  & 0.21         & 0.02 &  6.2     & 0.1 &-&-  G      \\
\ mll11 && 1.1    & 0.15      & 6.9     & 2.9 & 0.1*  & 330  & 200  & 0.16          & 0.57 &  2.4     & 0.1 & 21 & 0.3 \\ 
\ 2010kd R&0.101& 1.01   & 0.16  & 6.46    & 3.05& 0.09  &   -  &   -  &  -           &   -  &    -     &     - &-&-      \\
\ ml12  && 1.04   & 0.45      & 6.87    & 3.  & 0.24  & 900  & 570  & 0.21         & 0.02 &  6.2     & 0.1 &-&-        \\
\ mll12 && 1.1    & 0.15      & 6.9     & 2.9 & 0.1*  & 330  & 200  & 0.16          & 0.57 &  2.4     & 0.1 & 21& 0.3 \\
\ PTF10hgi I&0.099& 7.98 & -    & 6.66    & 7.08& 1.08  &  -   &   -  &   -          &   -  &   -      &      -&-&-  \\
\ ml13  && 7.2    & 0.4       & 7.4     & 3.1 & 1.    & 650  & 70   &   1.1        & 0.0016& 6.3     &   0.2 &-&-   \\
\ mll13 && 8.2    & 0.082     & 6.77    & 4.3 & 1.2  & 340  & 120  &  1.6         & 1.63  & 7.2     & 0.3   &60 & 0.6\\  
\ PS1-10bzj &0.649&0.98 & 0.07  & 8.21    & -   &  -    & -    &  -   &   -          &   -  &    -     &    - &-&-    \\
\ ml14   &&  1.2  & 0.44      & 8.13    & -   &  -    & 770  &  560  &   0.174      & 0.016& 5.6      &  0.1&-&-   \\
\ mll14  &&  0.97 & 0.17      & 7.8     & 2.9 & 0.1   & 330  &  210  &   0.16       & 0.58 & 2.6      &  0.1&21&0.4 \\
\ PTF10heh II&0.338& 2.   & -    & 3.58    & 2.4 &  0.15 &  -   &  -   &    -         &   -   &    -     &   - &- &-       \\
\ ml15   && 1.8   & 0.26      & 3.7     & 3.35& 0.42  & 540  & 190  & 0.29         & 0.0028& 2.      & 0.1         \\
\ mll15  && 2.    & 0.0067    & 3.32    & 3.0 & 0.3  & 280  & 90   & 10         & 0.23  & 5.8    & 0.3  &  3.7& 0.42\\   
\ PTF10qaf II&0.284& 1.8  & -    & 0.81    & 4.55& 1.47  & -    &  -   &     -        &    -  &  -      &  - &   - & -    \\
\ ml16    && 1.4  & 0.05      & 0.95    & 3.  & 1.35  & 790  & 120  & 0.9          & 0.013 &  2.     & 0.3&   - & - \\
\ mll16   && 1.7  & 0.003     & 0.88    & 3.  & 1.77 & 210  & 90   & 100          & 0.092 &  3.9    & 0.9& 3.4 & 0.1 \\
\ qaf SN loc&0.284& 2.7& -    & 2.36    & 2.  & 0.6   &  -   &  -   &     -        &   -   &   -     &   -&   - & -\\
\ ml17    && 3.1  &0.1        & 2.2     & 2.96& 0.6   &700   & 220  & 0.39         & 0.02  & 5.0     & 0.2& -   &-\\
\ mll17   && 2.7  & 0.016     & 2.47    & 2.94& 0.75 &240   & 130  & 0.4          & 0.026 & 2.7     & 0.2& 5.9 & 0.02 \\
\ PTF10vqv I&0.452& 1.3  & 0.09 & 6.33    & 1.87 & 0.8  &  -   &  -   &  -           &   -   &   -     &    -&-&-   \\
\ ml18    && 1.36 & 0.4       & 6.89    & 3.2  & 0.62 & 760  & 110  & 0.86         & 0.0013& 2.      & 0.3        \\
\ mll18   && 1.24 & 0.097     & 6.8     & 3.2  & 0.59 & 346  & 260  & 7.           & 64.5  & 0.9     & 0.1 &5.  & 0.1    \\
\ SN2011ke I&0.143& 0.98 & 0.113  & 5.67    & 2.9  & $<$0.033 & -&  -   &   -          &    -  &    -    &    -        \\
\ ml19    &&0.98    & 0.39    & 6.2     & 3.2  & 0.4  & 730  & 220  & 0.43         & 0.0028& 2.      & $<$0.2     \\
\ mll19   &&1.3     &0.09     & 6.3     & 3.29 & 0.07& 350  & 300  & 1           & 14.3  & 0.95    & 0.01  &5  & 0.1 \\   
\ tadpole &0.143& 5.06  & -   & 4.67    & 3.58 & $<$0.04& -  &   -  &    -         &    -  &    -    &     -&-&-  \\
\ ml20    && 4.9   & 0.34     & 5.1     & 3.2  &  0.57 &500  & 200  & 0.176        & 0.0036 &  5.    & $<$0.1     \\
\ mll20   && 5.2   & 0.07     & 4.7     & 3.78 &  0.12&340  & 120  & 0.16        & 1.37   &  7.4   & 0.04  & 40   & 0.6\\ 
\ SN2011kf I&0.245& 1.85  & -   & 6.1     & 3.5  & $<$0.38&-   &  -   &   -          &    -   &  -     &    -&-&- \\
\ ml21    && 2.    & 0.34     & 6.1     & 3.2  & 0.39  & 760 & 110  & 0.87         & 0.0016 &  2     & 0.1&-&- \\
\ mll21   && 1.98  & 0.017    & 6.0     & 2.94 & 0.42 & 250 & 100  & 2.           & 0.084  &  5.    & 0.3 & 5 & 0.1   \\
\ PS1-11ap R&0.524&3.13   &-    &4.8      & -    &   -   &  -  &  -   &   -          &    -   &  -     &   - &-&- \\
\ ml22    && 3.9   &0.2       &4.7      & 3.   & 0.22 &740  & 230  & 0.4          & 0.012  & 5.1    & 0.4 &-&-\\
\ mll22   && 3.4   & 0.012    & 4.88    & 3.   & 0.68 &220  & 90   & 100          & 0.1    & 6.5    & 0.3 & 5 & 0.1 \\
\ PTF11dsf II&0.385& 2.68  & -   & 4.9     & 2.9  & $<$0.15& -  &  -   &  -           &   -    &   -    &  - &-&- \\
\ ml23    && 2.4   & 0.26     & 5.2     & 3.46 & 0.3   & 540 & 190  &  0.24        & 0.002  & 8.     & 0.1&-&- \\ 
\ mll23     && 2.5  & 0.068   & 5.0     & 3.3  & 0.32 & 350& 100  & 25.        & 14.0    &  0.8   & 0.04  & 10. & 0.1  \\
\ SN2012il I&0.175&1.27   & 0.13 & 7.7     & 3.0  &$<$0.1 & -   &  -   &   -          &   -    &    -   &   - &-&-\\
\ ml24    &&1.24   & 0.43     & 8.1     & 3.   & 0.15 & 770 & 560  & 0.18         & 0.016  & 5.6    & 0.05&-&- \\
\ mll24   &&1.11   & 0.16     & 7.44    & 2.93 & 0.15 &330  & 200  & 0.16         & 0.57   & 2.6    & 0.1&21 &0.3\\
\ PTF12dam R&0.107&1.54   & 0.08 & 7.8     & 3.5  & 0.19  &  -  &  -   &   -          &   -    &  -     &  - &-&- \\  
\ mll25   && 1.56  & 0.1      & 7.3     & 3.3  & 0.4  & 350 & 300  & 1            & 14.3   & 1.1    & 0.06  &5  & 0.1   \\
\ SSS120810 I&0.156&2.7   &  -   & 2.88    & 3.   & $<0.24$  & - & -   &   -          & -      &   -    &  -&-&-   \\
\ ml26     && 2.7   & 0.11     & 2.4     & 3    & 0.4  & 730 &  220 & 0.39         & 0.02   & 5.3    & 0.05&-&-  \\
\ mll26    && 2.8  & 0.018     & 2.8    & 2.94  & 0.38 & 240 & 130  & 0.4          & 0.026  & 2.8    & 0.1 & 6.1 & 0.02 \\ \hline

\end{tabular}}

$^1$ in \erg

\end{table*}

\begin{table*}
\centering
\caption{Modelling  Sanders et al (2008) SN  host galaxy spectra. (Line ratios to \Hb=1)}
\tiny{
\begin{tabular}{lccccccccccccccccc} \hline  \hline
\  &z         & [OII]  & [OIII] & [OIII]  & \Ha & [NII] & [SII]&\Vs  & \n0  & $D$    & \Hb & O/H  & N/H &S/H& \Ts & $U$   \\ 
\  &          & 3727+  & 4363  & 5007+   &6563 & 6584+  & 6737+& \kms & \cm3 & 10$^{18}$ cm & flux$^1$ &10$^{-4}$ & 10$^{-4}$ &10$^{-4}$&10$^4$K & -  \\ \hline
\ {\bf Type}:IIb&&&&&&&&&&&&&&&\\
\ 2007ea & 0.04&4.15   &-      &3.3      &3.0  &0.65    &0.74  &    - &  -   &   -  &   -  &   - &   -  &   - & -  & -\\  
\ msnd1  &     &4.1    &-      &3.23     &2.95 &0.68    &0.8   & 300  & 120  & 0.5   &0.02 &6.6  &0.23  &0.05 &6.  & 0.025\\
\ 2010am & 0.02&2.97   &-      &3.94     &3.0  &0.64    &1.14  &  -   & -    & -     &   - & -   &  -   &  -   &  -& - \\
\ msnd2  &     & 3.0   &-      &4.0      &2.9  &0.66    &1.18  & 300  &120   & 0.5   &0.027&5.6  &0.26  &0.07  &6.2&0.04\\  
\ 2010cn &0.026& 3.0   &-      &3.94     &2.7  &0.27+   &0.68  &  -   & -    &  -    &  -  &  -  &  -   &  -   &  -& -  \\
\ msnd3  &     & 3.0   &-      &4.0      &2.9  &0.53    &0.69  & 300  &120   & 0.5   &0.028&5.2  &0.2   &0.04  &6.2&0.04\\
\ {\bf Type}:Ib&&&&&&&&&&&&&&&\\
\ 2006ja &0.077& 4.5   &-      &3.1      &3.1  &2.45    &1.02  &  -   & -    &  -    &  -  &  -  &    - &  -   &  -&  - \\
\ msnd4  &     & 4.3   &-      &3.1      &3.0  &2.4     &0.9   & 300  &100   &0.5    &0.014&6.6  &0.8   &0.05  &6.3&0.02 \\
\ 2007az &0.035& 2.15  &-      &5.26     &2.6  &0.18    &0.52  &  -   &  -   &   -   &  -  &  -  &  -   &  -   &  -&  -  \\
\ msnd5  &     & 2.2   &-      &5.16     &2.9  &0.21    &0.53  & 200  &100   &0.5    &0.027&6.2  &0.13  &0.03  &6.1&0.07 \\
\ 2007gg &0.038& 3.1   &-      &3.4      &3.5  &0.8+    &1.9   &  -   &   -  &   -   &   - &   - &   -  &   -  &  -&   - \\
\ msnd6  &     & 3.1   &-      &3.5      &3.0  &1.      &1.7   &300   &120   &0.5    &0.024&5.6  &0.4   &0.1   &6.2&0.033\\
\ 2008gc &0.049&2.9    &-      &3.0      &3.2  &0.56    &0.89  &  -   &   -  &   -   &   - &   - &   -  &    - &   -&  - \\
\ msnd7  &     & 3.1   &-      &3.1      &3.0  &0.58    &0.85  &320   &120   &0.6    &0.025&5.0  &0.2   &0.05  &6.0 &0.033\\
\ LSQ11JW&0.02 & 3.0   &-      &2.0      &2.6  &0.44+   &0.4+  &-     &   -  &    -  &   - &   - &    - &    - &  -&   -  \\
\ msnd8  &     & 3.1   &-      &2.0      &2.9  &0.5     &0.7   &300   &100   &0.6    &0.032& 6.0 &0.2   &0.05  &4.9&0.032 \\
\ PTF09dfk&0.016&1.97  &-      &4.2      &3.0  &0.53    &1.08  & -    &  -   &    -  &    -&   - &    - &    - &  -&   -  \\
\ msnd9  &     & 2.0   &-      &4.26     &3.0  &0.51    &0.96  &200   &100   &0.5    &0.027&6.2  &0.34  &0.06  &5.7&0.07 \\
\ {\bf Type:Ic}&&&&&&&&&&&&&&&\\
\ 2004ib &0.056&11.5   &-      &2.87     &3.0  &0.7     &1.12  &-     &  -   &  -    &  -  &   - &    - &   -  &  -&   -  \\
\ msnd10 &     &11.6   &-      &2.8      &3.1  &0.75    &1.27  &370   &130   &0.1    &0.007&7.0  &0.1   &0.04  & 15.&0.006\\
\ 2006ir &0.021&2.25   &-      &3.33     &3.0  &0.42    &0.72  & -    &   -  &  -    &  -  &   - &   -  &   -  &  - &  -  \\
\ msnd11 &     &2.3    &-      &3.4      &2.9  &0.42    &0.78  & 200  &100   &0.5    &0.023& 6.2 &0.25  &0.05  &5.4 &0.05 \\
\ 2007db &0.048&7.3    &-      &2.9      &3.0  &0.79    &1.27  & -    &   -  &  -    &   - &   - &   -  &   -  &  -  &  - \\
\ msnd12 &     &7.2    &-      &2.98     &3.0  &0.8     &0.92  &350   &130   &0.1    &0.01 & 6.2 &0.15  &0.04  &9.0  &0.01\\
\ 2007hl &0.056&8.84   &-      &1.2      &3.0  &1.1     &1.1   &  -   &   -  &   -   &   - &   - &   -  &   -  &   - &  - \\
\ msnd13 &     &8.3    &-      &1.2      &3.2  &0.8     &1.7   &380   &110   &0.3    &0.006& 8.0 &0.1   &0.03  &14.&0.0016\\
\ 2008ao &0.015&1.42   &-      &0.47     &3.0  &1.34    &0.36  &  -   &   -  &   -   &   - &  -  &   -  &   -  &   -  & - \\
\ msnd14 &     &1.7    &-      &0.47     &3.0  &1.0     &0.84  &320   &150   &1.5    &0.05 & 6.0 &0.5   &0.1   &3.5 &0.058\\
\ 2010Q  &0.054&2.12   &0.10   &6.4      &2.6  &0.22    &0.73  &  -   &   -  &   -   &   - &   - &   -  &   -  &   -& -   \\
\ msnd15 &     &2.2    &0.02   &6.47     &2.94 &0.22    &0.74  &200   &100   &0.5    &0.037&6.6  &0.15  &0.04  &6.6 &0.09 \\ 
\ msnd15-0 &    &2.2    &0.08   &6.7      &3.0  &0.21    &0.8   &200   &100   &2.4    &0.037&4.   &0.12  &0.03  & 6.3&0.055\\
\ 2011it &0.051&2.6    &-      &1.43     &3.6  &1.45    &1.65  &   -  &   -  &   -   &  -  &   - &   -  &   -  &  - &  -  \\
\ mnd16  &     &2.5    &-      &1.5      &3.0  &1.4     &1.3   &250   &100   &0.5    &0.02 & 6.3 &0.7   &0.1   &4.5&0.035\\  
\ PTF10bip&0.051&3.93  &-      &3.35     &3.4  &0.52    &1.18  &   -  &   -  &   -   &   - &   - &   -  &   -  &  - &  - \\
\ msnd17 &     & 3.91  & -     &3.48     & 3.0 &0.6     &1.04  & 300  &100   &0.5    & 0.017&6.0 &0.22  &0.06  &6.2 &0.026\\
\ {\bf Type:Ic-BL}&&&&&&&&&&&&&&&\\
\ 2006nx &0.137&6.0    &-      &4.7      &2.9  &0.55    &1.13  & -    &   -  &   -   &   - &  -  &   -  &   -  &  - &  -\\
\ msnd18  &     &6.0    &-     &4.75     &2.9  &0.53    &1.13  & 300  &100   &0.5    &0.019& 6.6 &0.14  &0.06  &7.9 &0.017\\
\ 2007ce  &0.046&1.19   &0.098 &7.5      &2.7  &0.99    &0.26  & -    &  -   &   -   &   - &  -  &   -  &   -  &  - &  -  \\
\ msnd19* &     &1.1    &0.096 &7.1      &3.0  &0.7     &0.2   & 150  & 60   & 0.98  &0.025& 8.0 &1.5   &0.3   &5.5 &0.08 \\
\ 2008iu & 0.130&0.86   &0.2   &8.3      &3.0  &0.3     &0.17+ &  -   &  -   &  -    &   - &   - &   -  &   -  &  - &  -  \\
\ msnd20*&     &2.     &0.67  &7.7      &3.5  &1.      &0.06  &800   &450   &0.003  & 0.005&5.  &0.4   &0.2   &0.8 &0.01 \\
\ 2010ah  &0.05 &3.8    &-     &2.4      &3.5  &0.33+   &0.13  &-     &  -   &   -   &   -  &  - &   -  &   -  &  - &  -  \\
\ msnd21  &    & 3.8    &-     &2.9      &3.0  &0.6     &0.8   &300   &120   &0.5    &0.002 &6.6 &0.23  &0.05  &5.4 &0.02\\
\ 2010ay  &0.067&2.33   &0.046 &7.52     &2.5  &0.33    &0.51  &  -   &  -   &  -   &    -  &   -&  -   &  -   &  - &  -\\
\ msnd22*  &    &2.0     &0.069 &6.1      &2.97 &0.3     &0.47  &200   &100   &2.4   & 0.034 &4.5 &0.2   & 0.2  & 6.3&0.055\\
\ {\bf Type:undet. Ibc}&&&&&&&&&&&&&&&\\
\ 1991R   &0.035&8.38   & -    &2.2       &3.0 & 0.96   &1.09   &  -  &   -  &   -  &    -  &  - &  -   &  -   &  -  &  -\\
\ msnd23  &     &8.3    &-     &2.2      &3.0  & 0.9    & 1.0   &360  &120   &0.1   & 0.006  &6.2 &0.15 &0.04  &6.3 &0.006\\
\ 2011gh  & 0.018&3.38  &-     &0.57     &3.0  &1.35    &0.88   &  -  &   -  &   -  &   -   &  -  &  -  &   -  &  - &  -  \\
\ msnd24  &     &3.4    &-     &0.56     &3.0  &1.39    &0.9    &200  &100   &1.5   &0.007  &6.4  &0.5  & 0.06 & 4.2&0.007\\
\hline
\end{tabular}}

$^1$ in \erg

\end{table*}

\begin{figure}
\centering
\includegraphics[width=9.0cm]{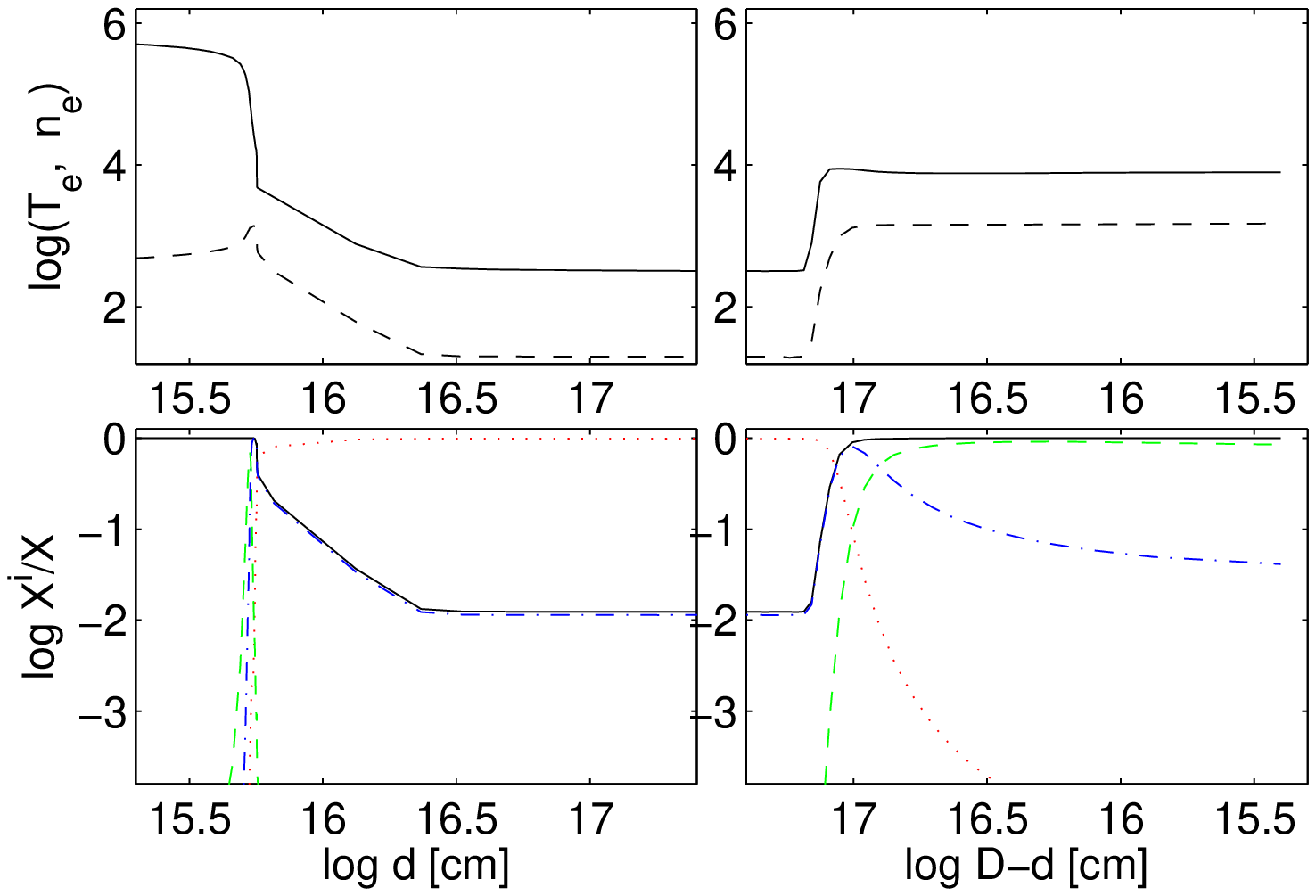}
\includegraphics[width=9.0cm]{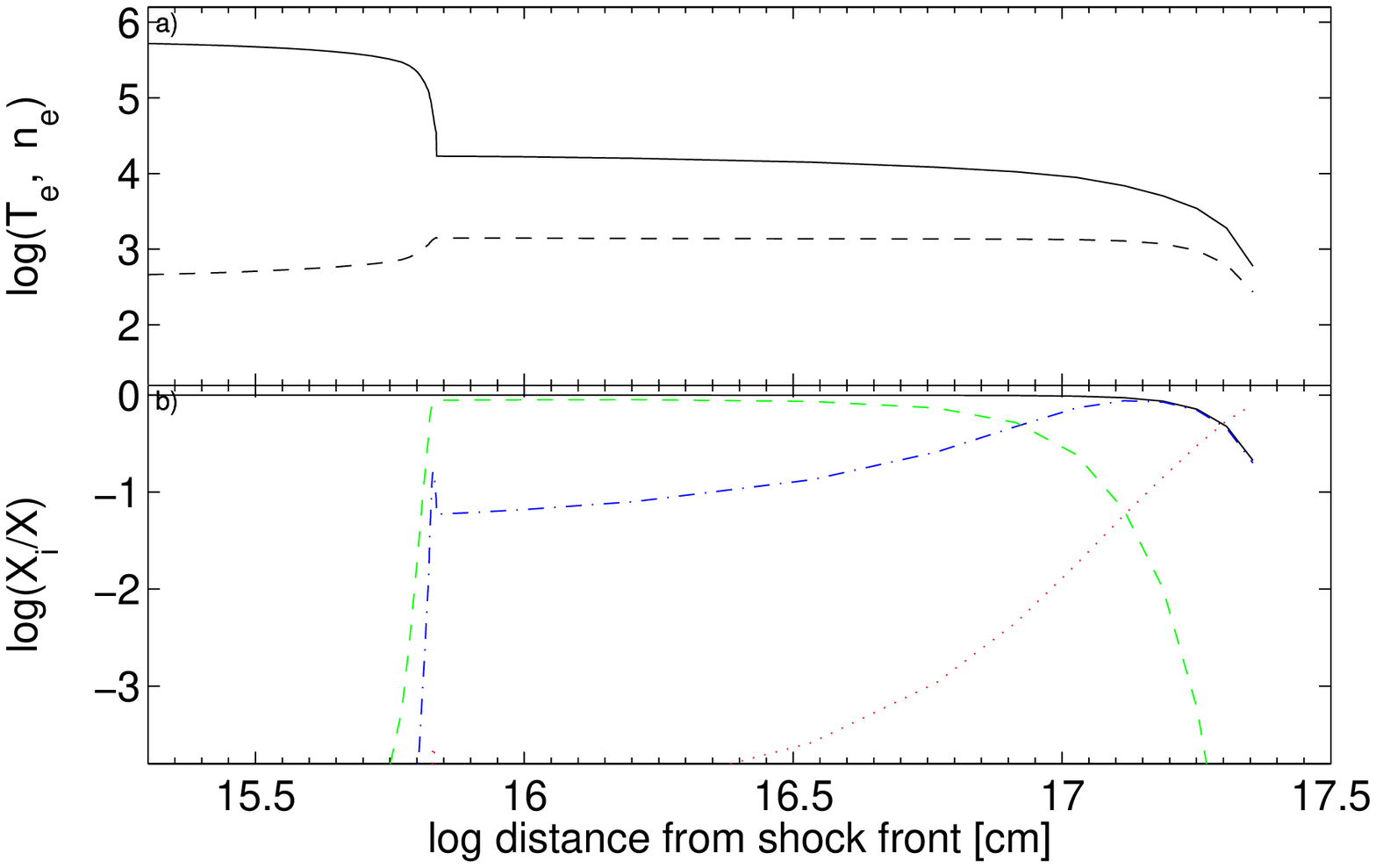}
\caption{ Top diagrams refer to the outflow case for model msnd15 (see text). 
The shock front is on the left and the radiation flux reaches the 
opposite (right) cloud edge.  
The bottom diagram refers to the inflow case for model msnd15-0. 
The shock front is on the left and the radiation flux reaches the same edge
at the left of the cloud. 
Top panels : the profile of the electron temperature (solid line) and the electron density
(dashed line) throughout the gas emitting clouds. Bottom panels ; the profiles of
O$^{++}$/O (dashed), O$^+$/O (dot-dashed, O$^0$/O (dotted) and H$^+$/H (solid).
}
\end{figure}

\begin{table*}
\centering
\caption{Modelling Modjaz et al (2008) Type Ic (broad-line) selected host galaxy central spectra}
\tiny{
\begin{tabular}{lccccccccccccccc} \hline  \hline
\                 &SN 2005kr&    &SN 2005ks&    &SN 2006nx&    &SN 2006qk & \\
\  z              &0.1345   &    &0.0987  &    &0.1370  &    &0.0584   &    \\
\ line ratios     & obs     &mM1 & obs    &mM2 & obs    &mM3 & obs     &mM4 \\ \hline
\ [OII]3727+      &1.67     &1.9 &2.97    &2.72&3.11    &3.2 & 3.54    &3.3 \\
\ \Hg             &0.37     &0.46&  -     & -  & -      & -  & 0.43    &0.46 \\
\ \Hb             & 1       &  1 &  1     & 1  &  1     & 1  &  1      & 1   \\
\ [OIII]5007+     &4.0      &4.1 &0.92    &0.88& 3.85   &3.3 & 0.33$^2$&0.6  \\
\ HeI 5876        &0.14     &0.15&  -     & -  &  -     & -  &  -      &  -  \\
\ [OI]6300+       & 0.13    &0.13&0.11    &0.29&   -    & -  &  -      &   -  \\
\ [NII]6563+      &0.38     &0.4 &1.4     &1.5 &   0.25 &0.47&  1.6    & 1.53 \\
\ \Ha             &3.46     & 3  & 3      & 3  & 3      & 3  & 3       & 3    \\
\ [SII]6717       &0.54     &0.50& 0.71   &0.63&  -     & -  & 0.75    & 0.65 \\
\ [SII]6731       &0.34     &0.9 & 0.52   &0.86&  -     & -  & 0.45    & 0.1  \\
\ \Vs (kms)       &   -      &300 &   -   &140 & -      &300 &  -      & 300  \\
\ \n0 (\cm3)      &  -      &220  &   -   &180 &  -     & 110&  -      & 90    \\
\ $D$ (10$^{16}$cm)& -      & 8   &   -   &100 &  -     & 5  &  -      & 20    \\
\ \Ts (10$^4$K)   &  -      & 5.4 &   -   &5.2 &  -     & 5.4&  -      & 4.2   \\
\ $U$             &  -      & 0.09&   -   &0.009& -     &0.041&  -     & 0.01  \\
\ O/H (10$^{-4}$)  &  -      & 6.2 &   -   &6.0 &  -     &6.2  &  -     & 4.6   \\
\ N/H (10$^{-4}$)  &  -      & 0.2 &   -   &0.6 &  -     & 0.2 &  -     & 0.6   \\
\ S/H (10$^{-4}$)  &  -      & 0.13&   -   &0.1 &  -     & 0.3 &  -     & 0.13  \\
\ \Hb (\erg)      & 8.2$^1$     &0.091& 69.6$^1$  &0.019&6.9$^1$     &0.027& 43.7$^1$ & 0.0073 \\ \hline
\end{tabular}}

$^1$ $\times 10^{-17}$

$^2$ only the 5007 component was observed

\end{table*}

\begin{table*}
\centering
\caption{Modelling Modjaz et al (2008) Type Ic (broad-line) host  observed spectra at SN positions}
\tiny{
\begin{tabular}{lccccccccccccccc} \hline  \hline
\                 &SN 1997ef&    &SN 2003jd&    &SN 2005kz&    &SN 2005nb& \\
\  z              &0.0117   &    &0.0188  &    &0.0270  &    &0.0238   &    \\
\ line ratios     & obs     &mM5 & obs    &mM6 & obs    &mM7 & obs     &mM8 \\ \hline
\ [OII]3727+      & 2.3     &2.5 &3.13    &3.3 & 0.51   &0.55& 3.11    &3.86\\
\ [NeIII]3869+    & -       & -  & 0.08    &0.087&-     & -  &   -     & -  \\
\ \Hg             & 0.42    &0.46& 0.43   &0.46& -      & -  & 0.5     &0.46 \\
\ \Hb             & 1       &  1 &  1     & 1  &  1     & 1  &  1      &  1  \\
\ [OIII]5007+     & 0.57    &0.57& 2.     &2.1 & *0.175 &0.18& 1.6     &1.56 \\
\ [OI]6300+       & -       & -  &0.057   &0.057&0.24   & 0.2&  -      &   -  \\
\ [NII]6563+      & 0.82    &0.87&0.53    &0.58&1.98    &1.0 &0.76     &0.79  \\
\ \Ha             & 3.      & 3. & 3.28   & 3. & 2.53   & 3.6& 3.      & 3.   \\
\ [SII]6717       & 0.52    &0.5 & 0.48   &0.4 &  -     & -  & 0.25    &0.27  \\
\ [SII]6731       & 0.4     &0.7 & 0.4    &0.5 &  -     & -  & 0.24    &0.36  \\
\ \Vs (kms)       &   -      &160 &   -   &160 & -      &170 &  -      &160   \\
\ \n0 (\cm3)      &  -      &130  &   -   &120 &  -     &220 &  -      &120    \\
\ $D$ (10$^{16}$cm)& -      &6    &   -   &4   &  -     &1000&  -      &3      \\
\ \Ts (10$^4$K)   &  -      &3.8  &   -   &5   &  -     &6   &  -      &4.9    \\
\ $U$             &  -      &0.015&   -   &0.015& -     &0.008&  -     &0.01   \\
\ O/H (10$^{-4}$  &  -      &6.6  &   -   &6.6 &  -     &5.6  &  -     & 6.6   \\
\ N/H (10$^{-4}$  &  -      &0.4  &   -   &0.25&  -     &1.5  &  -     &0.3    \\
\ S/H (10$^{-4}$  &  -      &0.1  &   -   &0.1 &  -     &0.1  &  -     &0.07   \\
\ \Hb (\erg)      &478$^1$ &0.013&229$^1$ &0.011&656$^1$&0.097 &  376$^1$ &0.008 \\ \hline
\end{tabular}}

$^1$ $\times 10^{-17}$

\end{table*}

\begin{table*}
\centering
\caption{Modelling  Kr\"{u}ler et al (2015) long GRB host galaxy spectra. (Line ratios to \Hb=1)}
\tiny{
\begin{tabular}{lcccccccccccccccc} \hline  \hline
\        &z     & [OII]&[NeIII]& H${\gamma}$  & [OIII] & \Ha & [NII]& \Vs  & \n0  & $D$ & \Hb & O/H   & N/H & Ne/H & \Ts & $U$   \\ 
\  &          & 3727+  & 3869   & 4360        & 5007+   &6563 & 6584  & \kms & \cm3 & 10$^{18}$cm & flux$^1$ &10$^{-4}$ & 10$^{-4}$ & 10$^{-4}$ &10$^4$K & -  \\ \hline 
\ 050416A & 0.654& 5.757& 0.4 & 0.623& 3.254  & 3.0   &  0.353&121\\
\ mk1    &      & 5.9  & 0.67  & 0.46 & 3.85   & 2.95 & 0.46 & 150 &100  & 0.4 & 0.009 & 6.6   & 1.3& 1.  &  7.0 & 0.008 \\
\ 050714B &2.44  & 9.5  & 0.4   & 1.7  & 2.3    & 3.   & 1.23 &96 &&&&&&&\\
\ mk2    &      & 9.1  & 0.5   & 0.46 & 2.44   & 3.   & 1.13 & 100 & 100 & 1. & 0.003 & 6.6 & 0.22 & 0.5 & 11 & 0.002 \\
\ 050915A & 2.527& 1.66 & -     & 0.28 & 2.17   & -   & 210. & -   &   - &   - &   -  &  -   &  -  &  -  &  -   &-      \\
\  mk3   &        & 2.0  & 0.1 & 0.46 & 2.15   & 2.9 & 0.34 & 230 & 120 & 2.  & 0.19 & 4.8  & 0.2 & 1.  & 3.6  &  0.2  \\  
\ 050824   & 0.828  &1.57  &0.72   & 0.40  & 9.8   &  3.0   &  0.13&123.6 \\
\ mk4    &        & 1.8  &0.77   &0.46  & 9.76   &3.   & 0.117& 200 & 150 &   8 &0.114 & 6.5  & 0.1 &  1  & 7.   &0.09\\
\ 06111A  & 0.76 & 1.67 & 0.33    & -    &5.67   & -   & 0.33 & 88.5&-    & -   & -    &   -  &  -  &  -  & -    &-\\
\ mk5    &       & 1.5  & 0.34    & -    &5.67    & -   & 0.27 & 70. & 120 & 1   &0.041 & 4.8  & 0.3 &0.7  &  5.  &0.047\\
\ 060814  & 1.9   & 3.537 &  0.0   & 0.0  & 4.67  &  3.0   &  0.0 &314 \\
\ mk6    &      & 3.3  &   0.2   & 0.46 & 4.6    & 2.9  & 0.3 &280  & 80   &0.4 &0.053& 6.6   & 0.2 & 0.7 & 4.5  & 0.07 \\
\ 061021  & 0.34 & 3.99  & 0.0 & 0.443&   4.06 & 3.0  & 0.158&62.7\\
\ mk7    &      & 3.7  & -       & 0.46 &4.15    &2.93  & 0.25&80   &100   &0.7 & 0.008& 5.1  & 0.1 & 1.  &6.2  & 0.008\\
\ 061202  & 2.25 & 3.83 &  0.0  &0.0 & 3.64  &  3.0   &  0.29&158.8 \\
\ mk8    &      & 3.8  &  -      &  -   &3.7     & 2.95 & 0.25 & 150& 100  & 1.3& 0.018&6.4   & 0.1 & 1.  & 6.2  &0.014 \\
\ 070110  & 2.35 &1.97  &  0.114 &  0.641 &  7.63 &   3.0  &   0.86&75.5 \\
\ mk9    &      & 1.6  & 0.21   & 0.46 & 7.55   & 2.9    & 0.83  &75 & 120  & 1  & 0.04 & 6.2  & 1.  & 0.4 &5.6  & 0.047\\
\ 070129  & 2.34 & 3.65 &  0.63   & 0.55 &  5.4   & 3.0  & 0.33 &185\\
\ mk10   &      & 3.9  & 0.7    & 0.45 & 5.2    & 3.1  & 0.57 & 200& 150  & 8  & 0.083 & 6.3 & 0.2 & 0.9 &  7.3 &0.033 \\
\ 070224  & 1.99 & 7.5  & 2.5     & 0.83 & 12.5   & -    & -    & 100& -    & -  &  -    &  -  &  -  &  -  &-      &-\\
\ mk11   &      & 8.   & 2.06    & 0.46 & 12.7   & 3.   & -    & 100& 100  & 0.2& 0.007 & 6.7 & 0.3 & 1.3 & 12.  & 0.008\\
\ 070306  & 1.5  & 2.17 & 0.228   & 0.793&  5.0   & 3.0  & 0.359&289\\
\ mk12   &      & 1.9  &0.26     & 0.45 & 5.14   & 3.3  & 0.64 & 280& 150  & 1  & 0.17  & 6.5 & 0.5 & 0.5 & 7.    & 0.05\\
\ 070318  &0.84  & 1.677&  0.233  &0.241 &  5.316 &  3.0 &  0.273&134\\
\ mk13   &      &1.9   & 0.27    &0.46  & 5.21   & 2.94 & 0.13 & 130& 100  & 1  & 0.05  & 6.2 & 0.12& 0.8 & 4.8   &0.062\\
\ 070328  &2.06  &1.1   & 0.45    & 0.09 & 2.45   &  -   &  -   & 224& -    & -  &  -    &  -  &  -  &  -  &  -    &  - \\
\ mk14   &      & 1.1  & 0.25    & 0.46 & 2.3    &  -   &  -   & 220& 190  & 7  & 0.78  & 4.  & 0.12& 2.  & 3.5   & 0.9\\
\ 070419B & 1.96 & 2.439&  0.0    &  0.0 &  3.030 & 3.0  & 0.0&208\\
\ mk15   &      & 2.22 & 0.11    & 0.46 & 3.0    & 2.9  & 0.15 & 210& 120  & 0.8&0.4    & 6.  & 0.12& 1.  & 3.5   & 0.5 \\
\ 070802  & 2.45 & 2.916&  0.246  &0.311 &  4.582 & 3.0 &  0.709&143\\
\ mk16   &      & 2.7  & 0.26    & 0.46 & 4.5    & 2.9  & 0.67 & 140& 100  & 1  & 0.037 & 6.2 & 0.4 & 0.7 &5.2    &0.03 \\
\ 071021  & 2.45 &  1.2  &0.11    &0.25  & 2.57   & 3.2  & 0.0  &240.7&-   & -  &  -    & -   &  -  &  -  & -     & -   \\
\ mk17   &      &  1.7  & 0.09   & 0.45 & 2.51   & 2.95 & 0.29 &240  & 250& 1. & 1.3   & 6.0 & 0.2 & 1.  & 3.3  & 1.2 \\ 
\ 080207  & 2.086& 2.028 & 0.73    & 1.2 & 4.836  & 3.0  & 0.68 &324&&&&&&\\ 
\ m18    &      & 1.8  & 0.5     & 0.46& 4.9    & 3.2  & 0.52 &320&80&1.1&0.084&6.6&0.5  & 1. & 9.7 & 0.04 \\
\ 080605  & 1.64 & 2.77 &  -      & -    & 5.0    & 3.   & 0.4  &194    &&&&&&\\
\ m19    &      & 2.76 & 0.26    & 0.46 & 5.2    & 2.94 & 0.34 & 200 & 80 & 1. & 0.037 & 6.6  & 0.22 & 0.7 & 5. & 0.05\\
\ 080804  & 2.2  & 1.25 & 0.97    &   0.0&5.8     & 3.0  & 0.0 &127.9 & -   & -   & -  &   -   &  -  &  -  &  -  &  -    \\ 
\ mk20   &      & 1.2  & 0.6     &   -  & 5.9    &3.    & -    & 130 &100  & 0.8& 0.086 & 6.6 & 0.6 & 1.5 & 6.3    & 0.08\\
\ 080805  & 1.5  &  2.6 &   0.0   &  0.0 &  2.485 & 3.0  & 0.226&136&&&&&\\
\ mk21   &      & 2.73 &  -      &  -   &  2.63  & 3.   & 0.4  & 130 & 100 & 1.2& 0.03  & 6.6 & 0.2 & 1.5 & 7.8   & 0.01 \\
\ 081210  & 2.06 & 1.   & 0.38    & 0.52 & 2.86   & -    &  -   & 282 & -   & -  &  -    &  -  &  -  &  -  &  -    &  -   \\
\ mk22   &      & 1.1  & 0.26    & 0.45 & 2.84   & -    &  -   & 280 & 100 & 10 & 0.2   & 6.8 & 0.6 & 1.7 & 4.5   & 0.15 \\
\ 081221  & 2.26 & 1.143&   0.0   &  0.0 &  1.884 &  3.0 &   0.409&225&&&&&\\
\ mk23   &      & 1.1  & -       &   -  & 1.92   & 3.27 & 0.46 & 230 & 180 & 25 & 0.29  & 6.  & 0.6 & 1.  & 4.8   & 0.07 \\
\ 090201  & 2.1  &2.099 &   0.203 &0.697 &  2.050 &  3.0 & 0.055&405.8&&&&&\\
\ mk24$_{OS}$&  & 1.   & 0.3     & 0.44 & 2.5    & 3.8  & 0.4  &400  &400  &0.2-20&0.003& 1.  & 0.2 & 0.3 & -     &  -   \\
\ mk46$_{0}$ && 1.5  & 0.37    & 0.45 & 1.9    & 3.3  & 0.17 & 400 &300  & 2  & 0.0072& 3.  & 0.1 & 0.4 & 9.    & 0.001\\ 
\ mk25$_{AGN}$&  & 1.2  &0.27    &0.46  & 2.3    & 3.   & 0.23 & 400 &250  &0.12& 0.66  & 6.  & 0.1 & 0.8 &4$^2$  & -    \\
\ 090407  & 1.448& 3.744&   0.518 & 0.192&  1.286 &  3.0 & 0.748&261.5&&&&&&\\
\ mk26  &       &3.7   &0.73     & 0.45 & 1.35   & 3.18 & 0.98 &260  & 120 & 1 & 0.015 & 5.  & 0.3 & 1.4 & 7.    & 0.008 \\
\ 090926B &1.24  &6.911&   0.648 &0.331 &  6.220 &  3.0 &   0.274&161&&&&&&\\
\ mk27   &      &6.5   & 0.7     & 0.46 & 6.1    & 2.9  & 0.39 & 160 & 110 & 0.8&0.024  & 6.6 & 0.1 & 0.7 &  9.   & 0.01 \\
\ 091018  & 0.97 &2.624 &   0.0   &  0.0 &  3.838 &  3.0 & 0.932&143&&&&&&\\
\ mk28    &     & 2.7  & -       & 0.46 & 3.86   & 2.9  & 0.9  & 140 & 120 & 1.4& 0.037 & 6.6 & 0.5 & 0.7 & 5.3   & 0.037\\  
\ 091127   &0.49 & 5.321&   0.209 & 0.466&   6.495&   3.0& 0.219&86.5&&&&&&\\
\ mk29   &      & 4.7  & 0.29    & 0.46 & 6.5    & 2.93 & 0.25 & 80  & 120 & 0.8& 0.011 & 6.6 & 0.1 & 0.4 & 7.  & 0.009\\
\ 100316D& 0.059&  2.83& 0.323  & 0.476&  5.300 &  3.0 & 0.175&-&&&&&&&\\
\ mk30*  &        & 2.8  & 0.36  & 0.46 & 5.5    & 2.94 & 0.33 & 140 & 100 & 1  & 0.03  & 6.6 & 0.2 & 0.8 & 5.6   & 0.003\\
\ 100418A& 0.623  & 3.907& 0.160 &0.482 &  3.450 &   3.0& 0.349&140&&&&&&&\\
\ mk31   &        & 3.7  & 0.26  & 0.46 & 3.5    & 2.94 & 0.47 & 140 & 100 & 1  & 0.018 & 6.6 & 0.2 & 0.7 & 5.6   & 0.015\\
\ 100424A& 2.47   &1.887 & 0.748 & .414 &  8.403 &  3.0 &   0.0&211&&&&&&&\\
\ mk32   &        & 1.9  & 0.6   & 0.46 & 8.4    & 2.9  & 0.25 & 210 & 150 & 1.5& 0.112 & 5.  & 0.2 & 0.7 & 6.    & 0.085 \\
\ 100615A& 1.4    &4.984 & 0.590 &0.245 &  8.883 &  3.0 & 0.375&117&&&&&&&\\
\ mk33   &        & 5.   & 0.66  & 0.46 &8.6     & 2.9  & 0.54 & 120 & 100 & 0.8& 0.014 & 6.8 & 0.2 & 0.7 & 8.4   & 0.013 \\
\ 100621A& 0.543  &2.047&  0.274 &0.425 &  4.608 &  3.0 & 0.179&199.5&&&&&&&\\
\ mk34   &       & 2.5  & 0.3    & 0.46 & 4.6    & 2.97 & 0.37 & 200  & 150&  2 & 0.075 & 5.  & 0.2 & 0.7 & 5.    & 0.05  \\
\ 100814A& 1.44  & 1.826&  0.555 &   0.0&  5.743 &  3.0 & 0.223&88.5&&&&&&\\
\ mk35   &        & 1.74& 0.43   & -    & 5.79   & 2.93 & 1.2  & 90   &150 & 0.8& 0.068 & 6.6 & 0.2 & 1.2 & 5.    & 0.057  \\
\ 100816A& 0.8  & 10.213&   0.851 &  0.0 &  0.880 &  3.0 &   1.064&266&&&&&&\\
\ mk36   &      & 8.7  & 0.8     & -    & 1.18   & 3.5  & 1.0  & 270  & 70 & 1.3& 0.003 & 7.4 & 0.2 & 0.5 & 28.   & 0.002\\
\ 110818A& 3.36 & 2.75 & 0.38    & -    & 9.2    & -    & -    & 215  & -  &  - &  -    &  -  &  -  &  -  &  -    &  -   \\
\ mk37   &      & 2.5  & 0.6     & -    & 9.1    & -    & -    & 220  & 130& 7  & 0.084 & 6.6 & 0.4 & 0.8 &  7.   & 0.06 \\
\ 110918A& 0.984& 2.29 & 0.245   &   0.0& 0.9    & 3.0  & 1.007&300.9&&&&&&\\
\ mk38   &      & 1.9  & 0.2     & 0.44 & 0.87   & 3.3  & 1.0  & 300  & 90 & 0.7& 0.03  & 6.6 & 0.8 & 0.9 & 7.5   & 0.0095\\ \hline
\end{tabular}}

$^1$ in \erg;
$^2$ in 10$^{10}$ photons cm$^{-2}$ s$^{-1}$ eV$^{-1}$ at the Lyman limit
\end{table*}
\begin{table*}
\centering
                   Table 6 - continued
\\
\tiny{
\begin{tabular}{lcccccccccccccccc} \hline  \hline
\        &z     & [OII]&[NeIII]& H${\gamma}$  & [OIII] & \Ha & [NII]& \Vs  & \n0  & $D$ & \Hb & O/H   & N/H & Ne/H & \Ts & $U$   \\ 
\  &          & 3727+  & 3869   & 4360        & 5007+   &6563 & 6584  & \kms & \cm3 & 10$^{18}$ cm &flux$^1$  &10$^{-4}$ & 10$^{-4}$ & 10$^{-4}$ &10$^4$K & -  \\ \hline 
\ 111209A& 0.677 &4.04 &  0.479 & 0.455&   5.604&   3.0& 0.246&96.4&&&&&&&\\
\  mk39  &      & 4.2  & 0.37    & 0.46 & 5.6    & 3.   & 0.23 & 120  & 100& 0.7& 0.015 & 6.6 & 0.1 & 0.6 & 6.4   & 0.014\\         
\ 120119A &  1.73 &   0.425 &   0.200 &   0.348 &   0.0   &   3.0   &  0.406&250 &&&&&\\
\ mk40     &        &  0.35   &   0.2   &  0.45   &   2     &   3.    & 0.4   &300 & 200 & 2. & 1.66& 6.6&1. & 1.4&5.&0.3\\ 
\  120422A & 0.283 &   6.171 &   0.223 &   0.481 &   2.505 &   3.0   &   0.447&77&&&&&&\\
\  mk41    &         & 6.2     &   0.5   &  0.46   & 2.5     &  3.     &  0.5& 80 &100  & 1. 7& 0.004&6.67&0.15&1.&7.&0.003\\
\  120624B & 2.197 &   3.434 &   0.437 &   0.0   &   7.644 &   3.0   &   0.786&188&&&&&\\
\ mk42     &         &   4.    &   0.49  &  0.46   &   7.8   &   2.9    & 0.74 &200&100&1.3&0.032& 6.6&0.3 &0.6& 7.&0.03\\ 
\  120714B & 0.398 &  3.92  &   0.44  &   0.2   &   3.96  &   3.0   &   0.240&94.4&&&&&\\
\ mk43     &         &  4.2   &  0.44   &   0.46  &   4.0   &   2.94  & 0.27   &100&120&1.& 0.013&6.3&0.1 &0.8&6.5&0.009\\ 
\  120722A & 0.959 &  3.067 &   0.166 &   0.327 &   3.827 &   3.0   &   0.383&141&&&&\\
\ mk44     &         &  3.1   &   0.24  &   0.46  &   4.    &   3.    &  0.45 &140&100& 3.&0.024 & 6.3&0.2&0.5&6.5&0.016\\
\  120815A & 2.36  &  1.5   &   0.91  &   0.0   &   7.12  &   3.0   &   0.37 &82.7&&&\\
\ mk45     &         &  1.4   &   0.7   &   0.46  &   7.3   &   2.9   &  0.38 &80&100 & 1.&0.038 & 6.6&0.6&1.5&5.4&0.056\\
\  121024A & 2.3  &  1.552 &   0.0   &   0.0   &   6.228 &   3.0   &   0.305&213.2&&&&&&\\
\ mk46     &        &  1.53  &   0.24  &   0.46  &   5.8   &   3.    &   0.23 & 200&120&0.6&0.3&6.9&0.4&1.&4.&0.6\\
\  130427A & 0.34 &  4.253 &   0.0   &   0.06  &   1.537 &   3.0   &   0.3&106.6&&&&&\\
\ mk47     &        &  4.24  &   0.3   &   0.46  &   1.69  &   3.    & 0.28 & 100 & 100&1.37&0.007&6.6&0.1&1.&5.7&0.005\\
\  130925A & 0.3483 &  3.302&   0.143 &   0.601&   1.956 &   3.0   &   0.65 &126&&&&&\\
\ mk48     &        & 3.7     &  0.2    &  0.46  &   1.97  & 3.      & 0.67   &130&100&1.3&0.012&6.6&0.26&0.8&5.4&0.009\\
\  131103A &  0.596 &  1.7  &   0.234 &   0.28 &   4.511 &   3.0 &   0.155&200&&&&&\\
\ mk49     &          & 1.8  &   0.2   &   0.46  &   4.4   &   2.94&   0.15 &200& 130&1. & 0.23 & 6.6& 0.15& 1.&4. &0.28\\
\  131105A & 1.685 &  8.378 &   1.145 &   0.549 &   2.132 &   3.0   &   0.404&132&&&&&\\
\ mk50     &         & 8.1    & 1.19    & 0.46    &   2.14  &   3.1   & 0.5 &130&150& 1.5&0.009 &6.6& 0.1 & 1. &14.&0.003\\
\  131231A & 0.64 &  2.22   &   0.302 &   0.67  &   5.894 &   3.0   &   0.3  &92&&&&&&\\
\ mk51     &        &  2.5    &   0.35  &  0.46   &   5.83  &  2.94   &  0.35 & 100&100&1. & 0.024&6.6&0.25&0.8&5.6 &0.026\\
\  140301A & 1.41  &  2.92 &   0.0   &   0.386 &   0.953 &   3.0   &   1.032&280&&&&&&&\\ 
\ mk52     &         & 2.8   & 0.3     & 0.46    & 1.1     &  3.4    & 0.97 & 280&100&2.5&0.027& 6.6 &0.5&0.8&9. &0.0078\\  \hline
\end{tabular}}

$^1$ in \erg

\end{table*}

\begin{table*}
\centering
\caption{Modelling  Savaglio  et al (2012) long GRB host spectra. (Line ratios to \Hb=1)}
\tiny{
\begin{tabular}{lccccccccccccccccccc} \hline  \hline
\        &z     & [OII]&[NeIII]& \Hg   & [OIII] & \Ha & [NII]& [SII] & \Vs  & \n0  & $D$ & \Hb & O/H   & N/H & Ne/H& S/H & \Ts & $U$   \\ 
\  &          & 3727+  & 3869   & 4360 & 5007+   &6563 & 6584  &6717& $^1$ &$^2$  &$^3$  & $^4$ &$^5$  &$^5$  &$^5$&$^5$  &$^6$ & -  \\ \hline 
\ 980425$^7$ &0.0085&2.07&0.49& -      & 6.76   & 3    & 0.18 & 0.16 &  -  &  -   &   -  & -     & -    & -   &   - &-    &    -   &    -  \\  
\ ms0       &       & 2.0 &0.5 &  -    &6.74   & 3     & 0.18 &0.19  &110  & 120  & 0.4  & 0.024 &6.0   &0.14 &0.8 &  -&6.6&0.078  \\
\ 990712    & 0.434 &2.68 &0.41 &0.34 &6.22    & 3.4  &$<$0.75& -    &  -  &  -   &   -   &   -  &  -  &  -  &  -  &  -& - &- \\             
\ ms1       &       &2.67 &0.45 &0.46 &6.3     & 2.94 &0.62   & -    & 100 & 100  & 0.5   & 0.012 & 6.6&0.4  &0.8  & - &7.3 & 0.03 \\
\ 020903    &0.251  &1.9  & 0.44& 0.35&7.23    & 3.6  &0.08   & 0.2  &   - &  -   &   -   &   -   &-   &  -  &  -  &  - & - &  -   \\
\   ms2     &       &2.0  & 0.5 & 0.46 &7.2     & 2.9  &0.13   & 0.2 & 110 & 120  & 0.4   & 0.024 &6.0 &0.1  &0.8  &0.02&6.8 & 0.079 \\
\ 030329    &0.168  & 1.74&0.34 &0.41  &4.87    &2.76  &0.15   &-    &  -  &  -   &  -    &  -    &  -  &-   &  -   &  - & -
&- \\
\ ms3       &       & 1.76&0.33 &0.46  &4.85    &2.94  &0.16   &0.16 &110  &120   &0.4    &0.024  & 6.0 &0.14 &0.85 &0.02 &5.6 & 0.078 \\
\ 031203    &0.1055 & 0.74 &0.44&0.42  &8.6     &2.94  &0.16   &0.09 & -   &  -   & -     & -     & -   & -   & -   & -   &-
& -\\
\ ms4       &       & 0.94 &0.5 &0.46  &8.54    &2.91  &0.17   &0.09 & 180 &350   &0.3    & 0.36  & 5.0 &0.13 &0.5  &0.04 &6.4&1.5 \\
\ 060218    &0.0334 & 2.11 &0.37& 0.52 &6.08    &3.4   &0.20   &0.15 &  -  &  -   &  -    &  -    &  -  &  -  &  -   &-   &-&- \\
\ ms5       &       & 2.1  &0.39 &0.46 &6.2     &2.9   &0.24   &0.19 & 100 & 100  &0.5    & 0.016 & 6.6 & 0.2 &0.8   &0.02 &6.5& 0.06\\
\ 060505    & 0.2889& 3.6  & -   &0.28 & 2.27   &3.69  &0.7    &0.78 &  -  &  -   &  -    &   -   &  -  &  -  &  -   &  -   &-&-\\
\ ms6       &       & 3.5  &-    & 0.46 & 2.3   & 3.   & 0.75  & 0.73& 130 & 100  &  1.3  & 0.003 & 6.6 & 0.3 & 0.8 & 0.08 &6.4&0.009\\ \hline 
\end{tabular}}

 $^1$ in \kms; $^2$ in \cm3 ;$^3$ in 10$^{18}$ cm; $^4$ in erg; $^5$  10$^{-4}$; $^6$ in 10$^4$K; $^7$ reddening corrected;
\end{table*}

\subsection{LGRB host galaxy spectra from the Savaglio et al (2009) sample}

We have selected from the sample of LGRB host galaxy spectra presented by Savaglio et al (2009)
the objects showing enough lines to constrain the models (Table 7).
Comparing the results for the O/H  relative abundances obtained by Savaglio et al by different  methods
and by detailed modelling  confirms that significant differences  can lead to different conclusions about the LGRB 
host galaxy nature. 
As mentioned in Sect. 2 the main theoretical reason of the gap derives from the fact that the lines within a galaxy are emitted from gas
in  various physical conditions. 
Detailed modelling was  used to  explain the conditions in local galaxies where the spectra  account for many lines
in different ionization levels and corresponding to many elements.
We believe that the same models should be used for galaxies at higher redshifts even if at the state of the art the
lines are few.

\begin{figure}
\centering
\includegraphics[width=9.0cm]{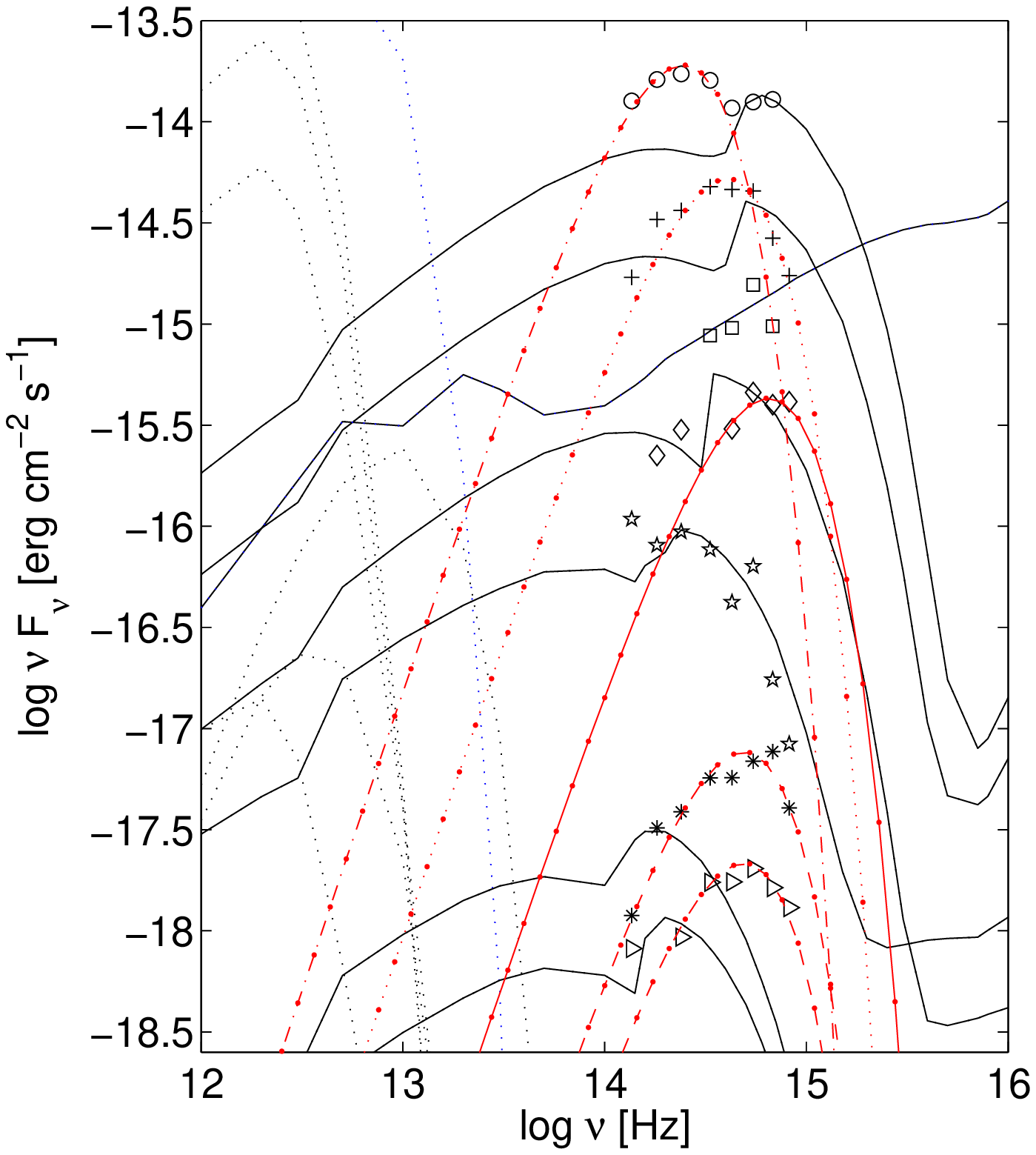}
\caption{
The continuum SEDs of LGRB with WR stars (Han et al).
Open circles:980703; plus:990712; open squares:020405; open diamonds:030329; open stars:
031203; asterisks:060218; open triangles: 060505b.  Solid lines : the bremsstrahlung calculated for each
galaxy by models mh1-mh8b. Dotted lines : dust reprocessed radiation. Black body flux calculated by T=8 10$^3$K 
(solid line + points), by T=6 10$^3$K
(dashed line + points), by T=5 10$^3$K (dotted line + points), T=3 10$^3$K (dash-dotted line + points) (see text). 
}
\end{figure}

\subsection{LGRB  host galaxies with different characteristics}

The modelling of  the host spectra of GRB 980425  presented by Sollerman et al (2005), GRB 991108 presented by
both  Castro-Tirado et al. (2001)  and Graham \& Fruchter (2013), GRB 010921, GRB 011121, 020819B, GRB 050824, 
GRB 050826 and GRB 070612A by Graham \& Fruchter (2013), GRB051022  by Levesque et al (2010), GRB 091127 by Vergani et al
(2011) and GRB 000210 by Piranomonte et al (2015)  is shown in Table 8. The models are given in Table 9.
Sollerman et al (2005) spectrum  was selected  because the survey  refer to low redshift 
galaxies (z$<$ 0.2) that host GRB.
They claim that they  are star-forming galaxies  (L$<$ \Lsol) with relatively low metallicity. 
 The spectra include the He I 5876 line which is also significant  in SN Type 1c hosts relatively to WR stars.
 Sollerman et al derived for the first time the SFR of the GRB 980425 host. They  have shown that this galaxy 
is not very metal-poor, and that a population study based on the broad-band photometry is consistent with a 
normal star forming galaxy with continuous star formation over 5 - 7 Gyrs, i.e., not with a starburst galaxy. 
A similar investigation of the spatially resolved H II region where the GRB occurred gives us an estimate 
of the GRB progenitor mass of $\geq$ 30 \msol  consistent with theoretical scenarios of SN 1998bw  which lead
to a very massive star (e.g. Iwamoto et al. 1998). 
We have found for 980425 \Vs=120\kms and  and \n0=150 \cm3, O/H close to solar and 
Ne/H solar, but low N/H and S/H similar to other galaxies at this z. 
The SB effective temperature is 6.5 10$^4$ K  higher than  Sollerman et al results.
The FWHM of the line profiles  observed by Castro-Tirado et al and reported in Table 3 for 991208 are  
broad (1200 \kms). The spectrum shows only a few lines, but the model is constrained by the high velocity shock.
O/H is slightly lower than solar, but Ne/H and N/H are solar.
In contrast, the spectrum  reported by Graham \& Fruchter shows \Vs similar to those which appear for other LGRB 
hosts in   the other surveys. For both models mGRB2a and mGRB2b $U$ is higher than generally found.
The  FWHM of 091127 host observed by  Vergani et al are rather narrow (50 \kms).
The [OII]/\Hb line ratio is unusually high. The modelling shows low \Vs and \n0 (60 \kms and 60 \cm3, respectively),
very thin clouds (0.03 pc) and O/H 1.12 solar, while N/H and S/H are low. 
Piranomonte et al present
VLT/X-Shooter spectra of the LGRB hosts at z$<$ 2.
We have selected GRB 000210 host galaxy spectrum from the Piranomonte et al (2015) sample
because  both the \Ha and \Hb lines were measured.
Our  modelling leads to "normal" parameters (mGRB11). 
For all the LGRB hosts presented in Table 8 O/H is close to solar.

\subsection{Han et al (2010) sample of LGRB hosts showing He, Ar and Fe lines }

Han et al present the spectra of LGRB hosts with richer data
 and determine  metallicities by direct methods.
We present in Table 10 the modelling of Han et al spectra and in Table 11
the models adopted to reproduce the line ratios.
Our results were obtained constraining the models on the basis of the He, Ar and Fe lines,
which are  seldom observed in the SN and GRB host spectra.
We have found in the host galaxies  shock velocities $\geq$ 200 \kms  and \n0 between 50 and 350 \cm3.  
Ar/H and Fe/H are by    
factors of 2 and 6 lower than solar, respectively. However, Fe/H   is based on  one only  object, 980703.
The O/H abundances calculated by the detailed modelling of the spectra are higher than those calculated by 
Han et al  adopting the direct method (Sect. 4.2). N/H and S/H are lower than solar up to a factor of 10.
To best reproduce the spectra of  980703 and 990712 we adopted He/H=0.13. Higher than solar (0.1) He/H
are found close to WR stars. In 980703 spectrum the HeII lines were not observed, but a high He/H=0.13 
was predicted by modelling. Helium is a strong coolant
and affects the calculation of the whole spectrum.
The large cloud geometrical thickness  reaching $\sim$30 pc were found by modelling the  060505b spectrum 
 which refers to the entire host.    
In most of the host galaxies $D$ $\sim$ 10 pc, while in
020405 and  060218 $D$ =  0.003 pc and 0.01 pc, respectively. 
These small clouds   have relatively high \n0. The   models were
calculated adopting   an inward motion of the clouds, towards the star-forming region. \Ts are low, 3 10$^4$ K,
but not exceptional.
Particularly low are the ionization parameters, indicating that  the emitting regions in these two
objects are  far from the SB stars.
 The SN associated with 020903 most probably occurred at several hundred parsecs from a bright,
relatively compact region responsible for WR and O star traces in the spectrum (Hammer et al 2006).
Accordingly, the calculated
 geometrical thickness of the emitting clouds  is  relatively large ($D$=3 pc).
We find for  031203 O/H solar and for 030329 O/H 0.34 solar. 
 A J-band image presented by Gal-Yam et al. (2004) for 031203 shows that the GRB occurred in the 
central regions of the host galaxy.

\begin{figure*}
\centering
\includegraphics[width=5.6cm]{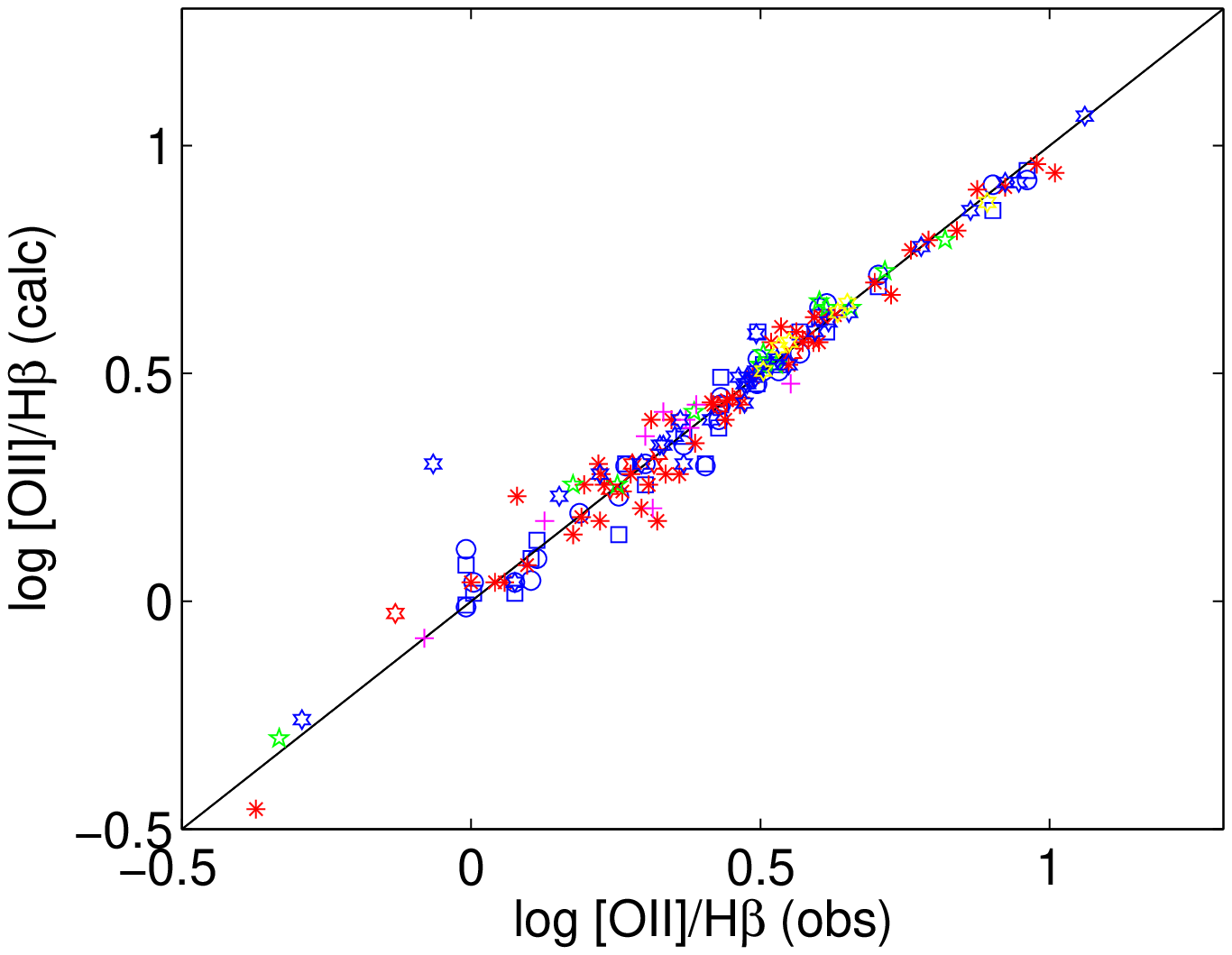}
\includegraphics[width=5.6cm]{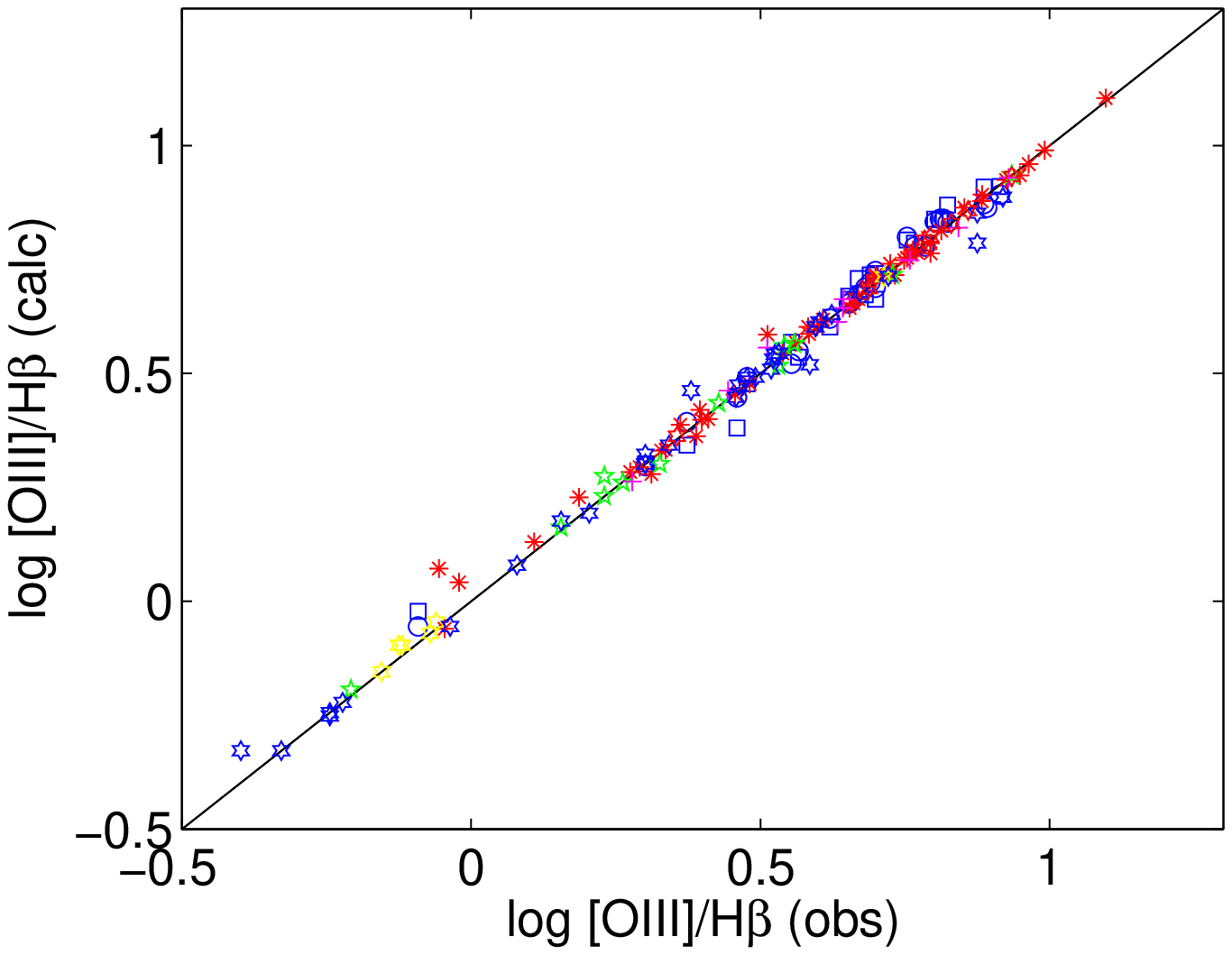}
\includegraphics[width=5.6cm]{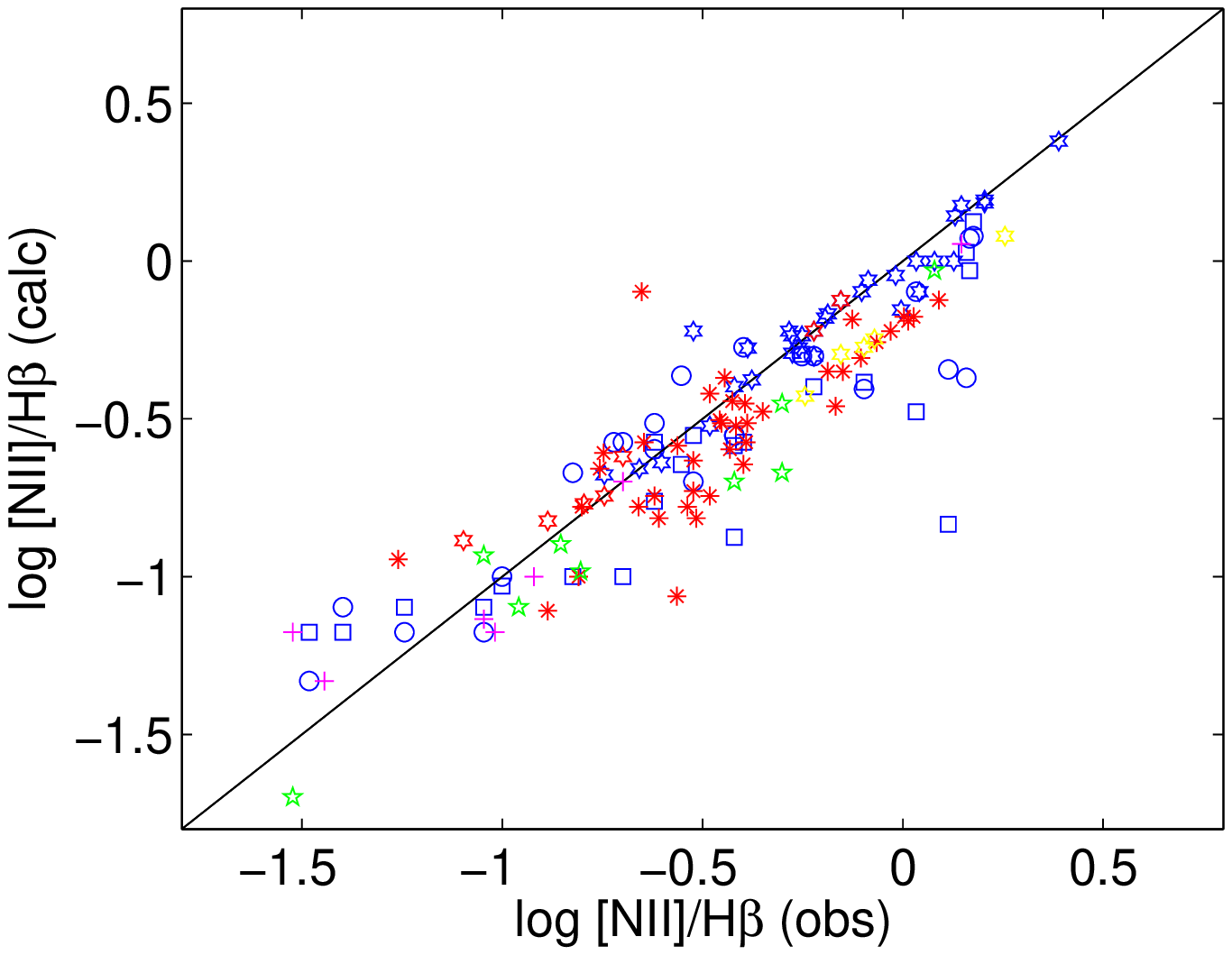}
\caption{Comparison of observed line ratios to \Hb  calculated by detailed modelling  with
 the data by Leloudas et al : blue circles (SB models), blue squares
(SD models); by Kr\"{u}hler et al : red asterisks;
by Sollerman et al ,  Graham \& Fruchter, Vergani et al, and Piranomonte et al : green stars;
by de Ugarte Postigo : yellow hexagrams;  by Han et al : magenta plus}
\end{figure*}

In Fig. 3 we present the modelling of the continuum SED of the Han et al host continuum emission. 
The observed  different object
fluxes are shifted along the Y-axis in order to compare   the SEDs in the same diagram.
The Y-axis scale refers only to 980703.
The data were not reddening corrected, therefore  the near-UV should be considered as lower limits.
For each galaxy we show the bremsstrahlung (black line)  calculated by the model which best fits the line 
spectrum (Table 11), the reprocessed radiation by dust calculated by a dust-to-gas ratio $d/g$=10$^{-15}$ by number
(dotted line) 
and the black body (bb) flux   referring to temperatures  selected phenomenologically   to
 best fit  the SED.
Reradiation by dust grains are shown only to indicate the peak frequency range.
The temperatures referring to the bb fluxes, which range between 3 and 8 10$^3$ K, 
are by a factor  $>$ 10 lower than those found for the  SB stars
by modelling  the line spectra. They represent an older stellar population throughout the hosts.
Except for 020405 and 031203 which can be reproduced by the bremsstrahlung alone, the other SEDs are well fitted by
the summed  bremsstrahlung  and bb fluxes. For 030329 two bb fluxes at different temperatures are present, while  for
990712, 060218 and 060505 the bremsstrahlung plays a negligible role.
The latter refers to the whole host.

\subsection{SGRB 130603B and SGRB 051221A host galaxies}

The spectra by FORS2 and X-shooter at VLT and ACAM at WHT were observed 
by de Ugarte Postigo et al (2014). They  claim that SGRB most probably derive from the merger of compact objects,
in particular for the short duration GRB 130603B, on the basis of the detection of "kilonova"-like signature
associated with {Swift}. The host galaxy is a perturbed spiral due to interaction with another galaxy
(de Ugarte Postigo et al). In  the spectrum taken by X-shooter the afterglow dominates the continuum, but
 the  lines  emitted from the host were used in their modelling. Therefore, we refer to line ratios and 
not to line fluxes. 
The FORS spectra   show the core,
the  arm and the opposite side of the galaxy. We   report  the X-shooter and FORS reddening corrected
spectra observed by de Ugarte Postigo in Table 12,  neglecting the GTC spectra because they do not include the
\Ha line. Each observed spectrum is followed by the best fitting model (mS1-mS4) in the next column.
The spectrum observed by Cucchiara et al (2013) for SGRB 130603B is also reported in Table 12 for comparison.

The models show shock velocities and pre-shock densities in the norm. However, to best reproduce all the line ratios
we adopted a magnetic field \B0 higher by a factor of 3 than  for the LGRB galaxies presented in the
previous sections. A higher \B0 prevents compression in the downstream gas, but its effect on the
line ratio results  is different than that obtained reducing the preshock density (see Contini 2009).
We find O/H near solar in all the positions in agreement with de Ugarte Postigo et al results,
and N/H and S/H lower than solar by  factors $\geq$ 2. S/H in
FORS (OT site) is nearly solar. The cloud geometrical thickness is relatively large  ($D$=1.8 pc) in
the X-shooter (OT site). $D$=0.3 pc  is used to fit the  FORS spectra, showing a larger cloud fragmentation.
The spectrum reported by Cucchiara et al (2013) is reproduced by
  model (mS5) similar to those used to fit the spectra  observed  by de Ugarte Postigo et al.
Star temperatures and ionization parameters referring to the SGRB are rather low ($\sim$ 3.5 10$^4$ K and $\sim$ 0.01, 
respectively).

Most of the spectra presented for SGRB surveys (e.g. Fong et al 2013, Berger et al 2005) do not contain
enough line to constrain the models.
We report in Table  12, last two columns, the spectrum observed by Soderberg et al (2006) for SGRB 051221a at z=0.546
and  model mS6.
The line ratios were reddening corrected adopting H${\gamma}$/\Hb=0.46.
The best fitting model shows 12+log(O/H)=8.8, in agreement with Soderberg et al who calculated 8.7  by the R32 method
(upper branch). We have found \Ts=8.2 10$^4$K and $U$=0.007.
\Ts are higher than for SGRB 130603B, where  \Ts  is rather low ($\sim$ 3.5 10$^4$ K). $U$, however, are similar.
For both SGRB 130603B and SGRB 051221a  the ionization parameters
are relatively low, roughly indicating that the  host observed positions  are far from the star-forming region.
Compared with LGRB hosts  reported  in Table 8,  we cannot find any significant difference 
in the physical conditions of the  host, except for a higher magnetic field in   130603B.

\begin{figure*}
\centering
\includegraphics[width=14.6cm]{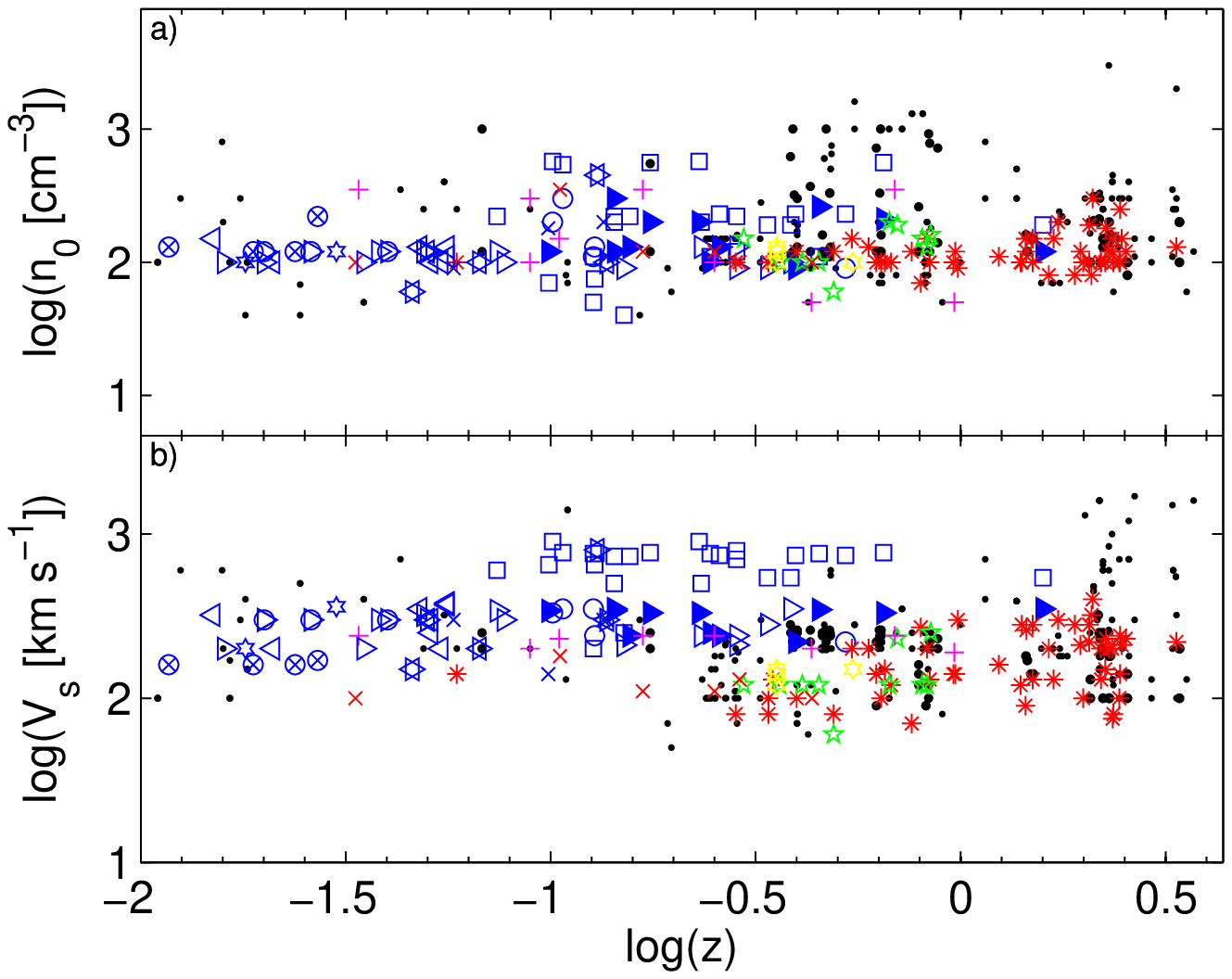}
\includegraphics[width=14.6cm]{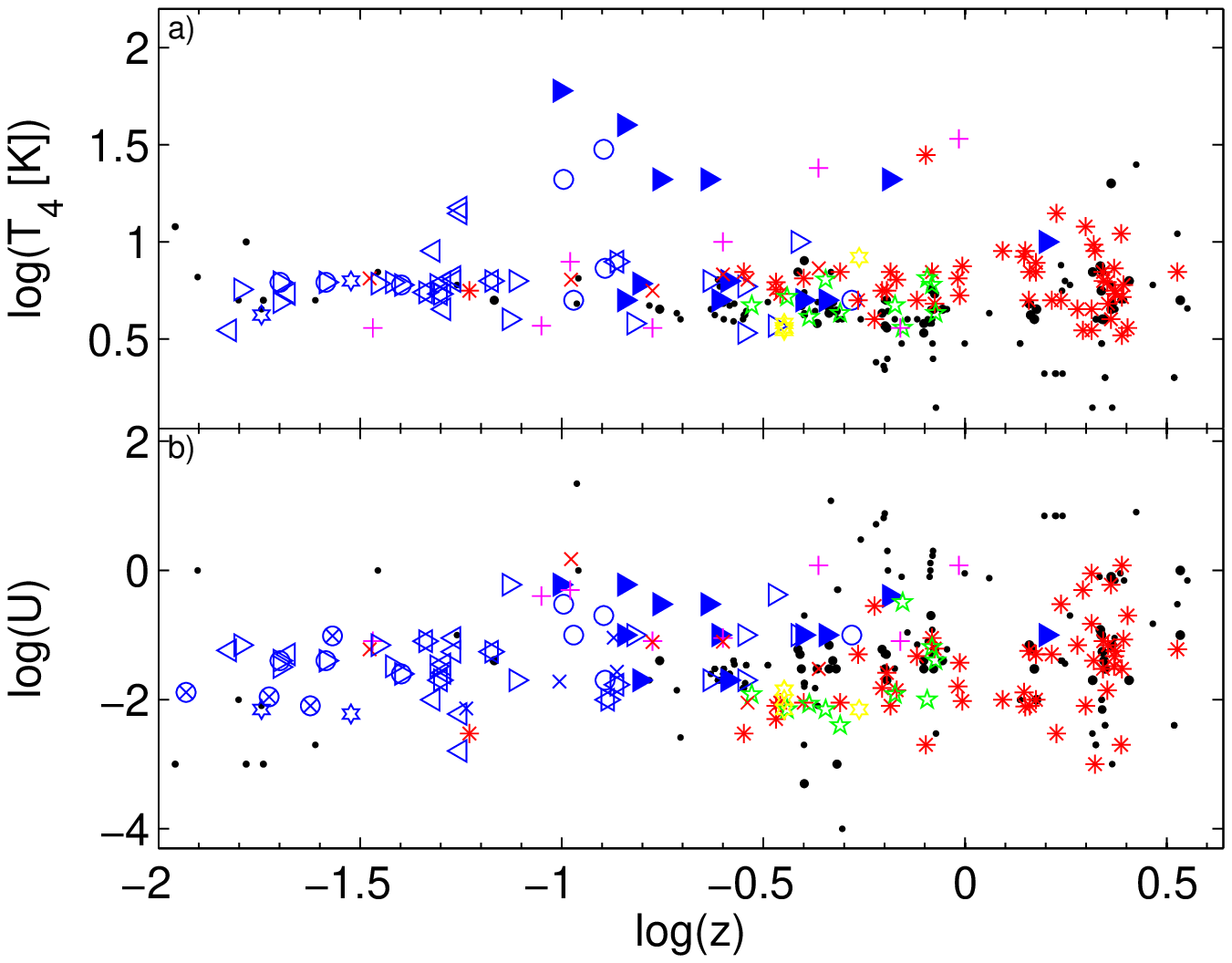}
\caption{ Top : \Vs and \n0 calculated  for the galaxy surveys as function of z.
Bottom : \Ts (in 10$^4$K) and $U$.  Symbols are  described in Table 13.}
\end{figure*}

\begin{figure*}
\centering
\includegraphics[width=14.6cm]{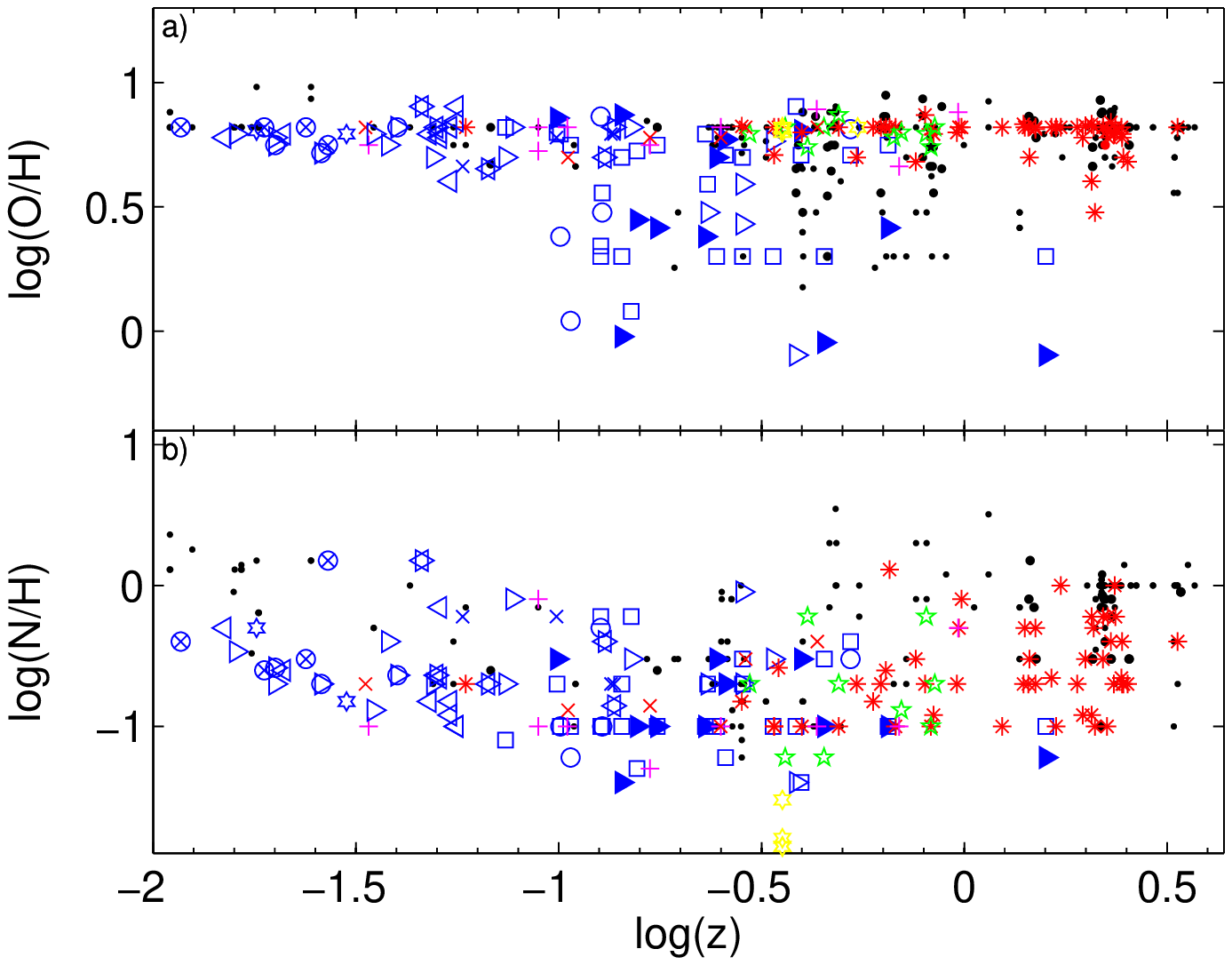}
\includegraphics[width=14.6cm]{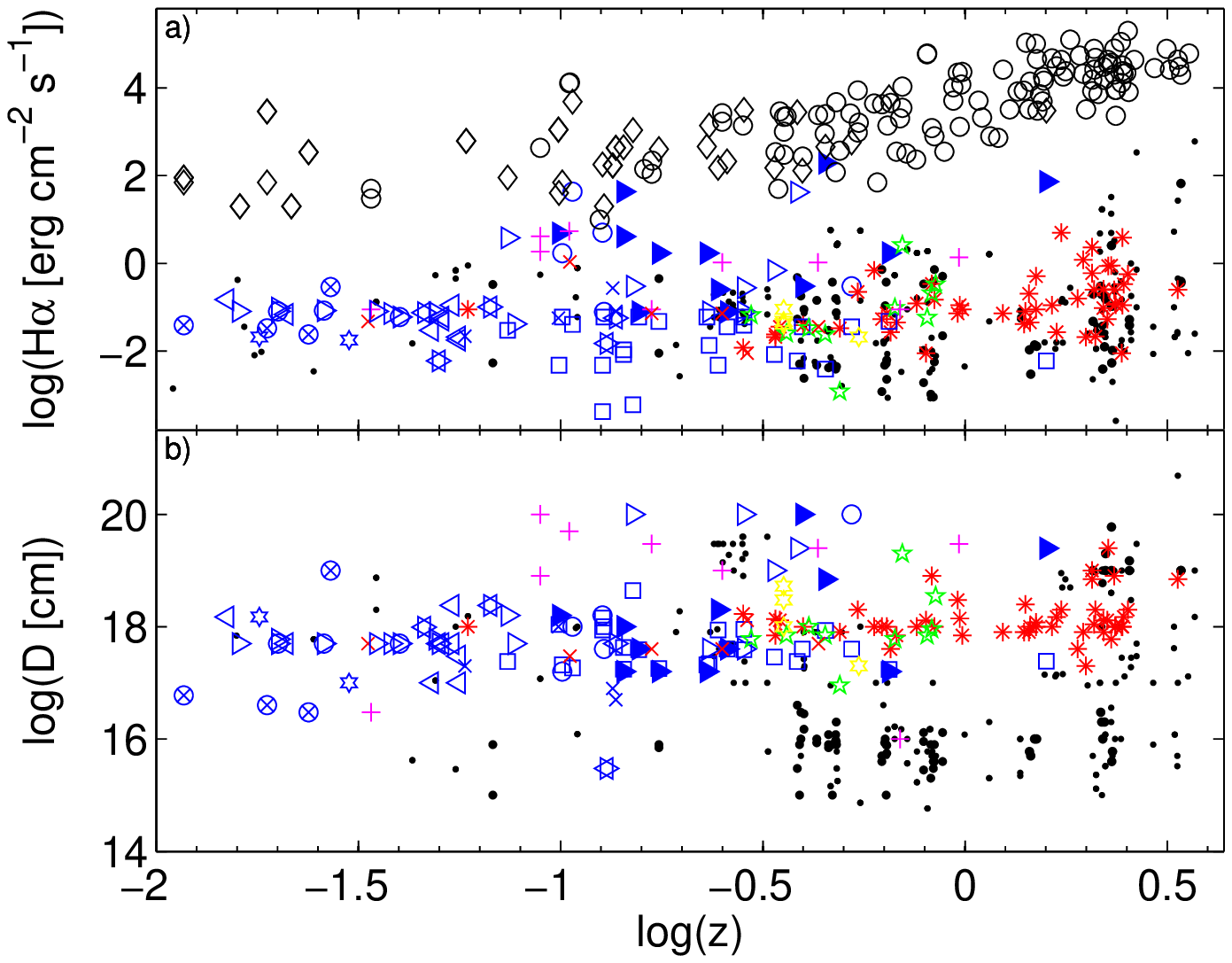}
\caption{Top : O/H  and N/H calculated for the galaxy surveys as function of z.
Bottom : \Ha and $D$. Symbols as in Fig. 5  are described in Table 13. 
Black open circles and open diamonds
: SFR referring to GRB and SN host galaxies, respectively.}
\end{figure*}

\begin{table*}
\centering
\caption{Modelling  LGRB hosts from different samples}
\tiny{
\begin{tabular}{lcccccccccccc} \hline  \hline
\             &z     & [OII]& [NeIII]& H${\gamma}$ & [OIII] & HeI  & \Ha & [NII]& [SII]   \\ 
\              &      &3727+ & 3689  &4360         & 5007+  &5876  &6563 & 6583 & 6726    \\ \hline
\ 980425$^1$ &0.0085&4.54  &0.64   &0.44         & 3.48   & 0.15 & 2.46& 0.38 & 0.89    \\
\ mGRB1        &      &4.4   &0.6    &0.46         & 3.63   & 0.18 & 2.95& 0.3  & 0.7     \\
\ 991208$^2$  & 0.7  &0.466 &   -   &  -          & 1.7    &  -   & -   &  -   &  -      \\ 
\ mGRB2a        &      &0.5   &   -   &   -         & 1.88   &  -   &  -  &  -   &  -      \\ 
\ 991208$^3$  &0.7  & 1.79  &-   & -   &2.12 & -      & 3.   & 0.157 & - \\
\ mGRB2b              &     & 1.8   & -  & -   & 2.0 &  -     & 3.   & 0.156 & - \\
\ 010921$^3$ & 0.451 &5.19  & -     & -           & 3.4    &  -   & 3.  & 0.14 &  -     \\
\ mGRB3       &       & 5.3  & -      & -          &  3.285  & -   & 2.95& 0.19 &  -     \\
\ 011121$^3$ & 0.362 & 4.   &  -     & -          & 1.83   &  -   & 3.  & 0.09 & -      \\
\ mGRB4       &       & 4.55 &  -     & -          & 1.82   &   -  & 2.96& 0.175& -       \\
\ 020819B$^3$& 0.411 & 2.43 &  -     &  -         & 0.62   &  -   & 3.  & 1.2  &  -     \\
\ mGRB5       &       &2.6   &   -    &   -        & 0.64   &  -   &2.97 & 1.4  &  -  \\
\ 050824$^3$ &  0.828&1.5   &   -    &  -         & 8.6    &  -   &3.   & 0.11 &  -  \\
\ mGRB6       &       & 1.8  &   -    &  -         &8.61    &  -   & 2.9 & 0.12 &  -   \\
\ 050826$^3$ & 0.296 & 3.2  &   -    &  -         &1.7.    &  -   & 2.97& 0.5  &  -   \\
\ mGRB7       &       &3.5   &   -    &  -         & 1.7    &  -   & 2.95& 0.53  & -    \\
\ 051022$^4$ & 0.806 &4.12  & 0.29   & 0.46       & 3.62   &  -   & -   & -     & -  \\
\ mGRB8       &       &4.4   & 0.34   & 0.46       & 3.67   &  -   & -   &  -    &  -  \\ 
\ 070612A$^3$ & 0.671&3.16   & -      & -          &1.43    &  -   & 3   & 0.03  & -    \\
\ mGRB9       &      &3.3    & -      & -          & 1.45   &  -   & 3   & 0.03  & -   \\
\ 091127$^5$ & 0.49 &6.6    & $<$1.4 &0.5         & 5.35   & -    &2.75 & $<$0.5 & 0.75 \\
\ mGRB10      &      &6.2    & -      &0.46        & 5.2    & -    & 3   & 0.32  & 0.06\\
\ 000210$^6$ & 0.846&3.37   & 0.136  &0.37        &2.68    & -    &2.8  & -     & -   \\
\ mGRB11      &      &3.3    & 0.16   & 0.46       & 2.72   & -    &2.94 & -     & -   \\ \hline

\end{tabular}}

$^1$ Sollerman et al(2005);
$^2$ Castro-Tirado et al (2001);
$^3$ Graham \& Fruchter (2013);
$^4$ Levesque et al (2010);
$^5$ Vergani et al (2011);
$^6$ Piranomonte et al (2015)

\end{table*}

\begin{table*}
\centering
\caption{The models  for LGRB  hosts in Table 8}
\tiny{
\begin{tabular}{lcccccccccccc} \hline  \hline
\             &\Vs & \n0 & $D$         & \Hb  & O/H     & N/H    & Ne/H      & \Ts     & $U$   \\  
\             &\kms&\cm3 & 10$^{18}$cm & \erg &10$^{-4}$&10$^{-4}$&10$^{-4}$& 10$^4$K & -     \\ \hline        
\  mGRB1      &120 &150  &0.65         &0.0186&6.3      & 0.09    & 1.      &6.5       &0.01    \\
\  mGRB2a      &1260&820  &0.182        &0.16  &6.2      &1.       & 1.       & 3        & 0.1    \\
\  mGRB2b     & 230 & 190 & 20.         & 0.86 & 6.3     &0.13     & 1.7      & 3.6     & 0.32   \\ 
\  mGRB3      &120 &100  &0.7          &0.008 &6.6      &0.06     &0.6       &6.4       &0.007   \\
\  mGRB4      &120 &100  &0.7          &0.0085&6.6      &0.06     &0.6       &5.2       &0.007   \\
\  mGRB5      &120 &100  &1.           &0.011 &5.5      &0.6      &0.6       &4.1       &0.0086  \\
\  mGRB6      &120 &130  &0.9          &0.069 &5.5      &0.1      &0.6       &6.        &0.065   \\
\  mGRB7      &120 &150  &0.6          &0.022 &6.2      &0.2      &0.6       &4.7       &0.012   \\
\  mGRB8      &120 &150  &0.7          &0.0193&6.2      &0.6      &0.6       &6.5       &0.01    \\
\  mGRB9      & 120 &200  &0.6         & 0.032  &6.      &0.01    &0.6        & 4.7    & 0.012  \\
\  mGRB10     & 60  & 60  & 0.09       &0.0004  &7.4     &0.2     &0.4        & 4.3    & 0.004   \\
\  mGRB11     & 250 &160  & 3.5        & 0.11   & 6.6    &0.2     &0.6        & 4.3    & 0.04    \\ \hline
\end{tabular}}

\end{table*}

\begin{table*}
\centering
\caption{Modelling  Han et al. (2010) host spectra. (Line ratios to \Hb=1)}
\tiny{
\begin{tabular}{lcccccccccccccccc} \hline  \hline
\        &z     & [OII]& H${\gamma}$ & [OIII]&[FeIII]&HeII&[ArIV]& [OIII] & \Ha & [NII]& [SII]&[SII]   \\ 
\        &      &3727+ & 4360        &4363   &4658   &4686&4711+ & 5007+  &6563 & 6583 & 6716 &6731   \\ \hline
\  980703&0.966 & 3.57 & 0.46        & -     &0.02   &-    &0.15 & 2.78    &  - &   -  &   -  &  -    \\
\ mh1    &      & 3.   & 0.42        & -     & 0.029 & -   &0.12+& 2.9     &  -  &   -  &   -  &  -    \\     
\ 990712 &0.433 & 2.06 & 0.43        & 0.08  & -     & 1.1 &  -  & 5.73    &3.75 & $<$0.036&-  &  -     \\
\ mh2    &      & 1.6  & 0.45        & 0.083 & -     & 1.2 &0.02 &5.6      &3.2  &0.07   &  -  &  -     \\
\ 020405 &0.691 & 2.15 & 0.35        & -     &  -    &0.009&0.015&4.39     & -   &  -    &  -  &  -     \\
\ mh3$^1$&      & 2.6  & 0.46        & -     & -     &0.01 &0.011&4.4      & 2.9 &  -     &  - &   -    \\
\ 020903 &0.251 & 1.34 & 0.46        & 0.064 & -     &0.045&0.14 & 6.96    & 3.03& 0.09   & 0.25&0.15   \\
\ mh4  &        & 1.5  & 0.45        & 0.04  & -     &0.03 &0.002 &6.6     & 3.2 & 0.11   & 0.14& 0.16  \\
\ 030329&0.168  & 2.   & 0.45        & 0.072 &-      &-    &-     &4.3     & 3.16& 0.03   & 0.39& 0.2  \\
\ mh5$^1$ &     & 2.3  & 0.46        & 0.05  & -     & -   &-     &4.1     &2.9   &0.1    & 0.02 &0.04  \\
\ 031203&0.105  & 0.83 & 0.44        & 0.07  & -     &0.001&0.008 &8.49    &3.2   &0.096  & 0.11 & 0.08 \\
\ mh6   &       & 0.83 & 0.46        & 0.05  & -     & 0.01&0.002 & 8.5    &3.11  &0.1    & 0.2  &0.2   \\  
\ 060218&0.034  & 2.45 & 0.46        & 0.05  & -     &0.009& 0.05 & 4.4    &2.89  &0.12   & 0.17 &0.13  \\
\ mh7$^1$&      & 2.7  & 0.46        & 0.05  & -     &0.01 & 0.02 & 4.6    &2.89  &0.15   & 0.04 &0.08  \\
\ 060505a&0.089 & 2.35 & 0.43        & -     &  -    & -   & -    & 3.25   &3.5   & 0.2   & 0.39 & 0.31 \\
\ mh8a   &      & 2.5  &0.46         & -     &  -    & -   & -    & 3.6    &3.    & 0.3   & 0.5  & 0.6  \\
\ 060505b&0.089 & 2.39 &-            & -     &  -    & -   & -    & 1.9    & 5.6  & 1.4   & 1.25 & 0.66 \\
\ mh8b   &      & 2.4  &-            & -     &  -    &  -  & -    & 1.83   & 3.1  & 1.7   & 0.6  & 0.7  \\ \hline

\end{tabular}}
\end{table*}

\begin{table*}
\centering
\caption{The models for Han et al (2010) host galaxies in Table 10} 
\tiny{
\begin{tabular}{lcccccccccccccccc} \hline  \hline
\        & \Vs  & \n0  & $D$         & \Hb      & O/H      & N/H       & S/H      &Ar/H     &Fe/H     & \Ts & $U$   \\ 
\        & \kms & \cm3 & 10$^{18}$cm &\erg &10$^{-4}$ & 10$^{-4}$ & 10$^{-4}$&10$^{-4}$&10$^{-4}$&10$^4$K & -  \\ \hline 
\ mh1    &190   & 50   & 30         &0.46     &7.6       & 0.5       & 0.01       & 0.033      &0.005 &34   & 1.2  \\   
\ mh2    &200   & 50   & 25         &0.35     &7.8       & 0.1       & 0.01       & 0.033      & 0.005&24    &1.2  \\
\ mh3    &240   & 350  & 0.01       &0.03     &4.6       & 0.1       & 0.01       & 0.033      & 0.005&3.6   &0.08  \\
\ mh4    &240   & 100  & 10         &0.35     &6.6       & 0.1       &0.033       & 0.033      & 0.005&10    & 0.09 \\
\ mh5    &240   & 350  &30          &0.03     & 5.6      & 0.05     & 0.09        & 0.033      & 0.005 & 3.6  & 0.08  \\
\ mh6    &230   & 150  & 50         &1.8      & 6.6      & 0.1      & 0.028       & 0.033      & 0.005 & 7.9  & 0.5  \\
\ mh7    &240   & 350  & 0.03       &0.03     & 5.6      & 0.1      & 0.2         & 0.033      & 0.005 & 3.6  & 0.08 \\
\ mh8a   &200   & 300  & 8          & 1.39    & 5.3      & 0.1      & 0.04        & 0.033      & 0.005 & 3.7  & 0.4  \\
\ mh8b   &200   & 100  & 100        & 0.62    & 6.6      & 0.8      & 0.1         & 0.033      & 0.005 & 3.7  & 0.4  \\ \hline

\end{tabular}}
\end{table*}

\begin{table*}
\centering
\caption{Modelling  SGRB 130603B  at z=0.356 and SGRB 051221a at z=0.546 host galaxy spectra}
\tiny{
\begin{tabular}{lccccccccccccccc} \hline  \hline
\ line         &OT site$^1$&mS1    &OT site$^1$&mS2  &core$^1$&mS3&  arm$^1$ &mS4&obs$^2$ &mS5 & 051221a$^3$&mS6\\
\              &X-shooter  &       &FORS       &     &FORS&       &FORS  &       &        &    & ....   &\\ \hline
\ [OII]3727+   & 4.47      & 4.5   &3.4        & 3.6 &3.55&3.7    &4.3   & 4.3   &3.05& 3.2    & 7.8  & 7.5\\
\ H${\gamma}$  &  -        &0.46   &0.89       & 0.49&0.49&0.46   &[0.5] & 0.46  & -  &0.46    & 0.46 & 0.46\\
\ \Hb          &  1        &1      &1          &1    & 1  & 1     &1     &1      &1   &1       & 1    &1 \\
\ [OIII[5007+  &  0.87     & 0.9   &0.76       & 0.8 &0.75&0.8    & 0.85 & 0.85  &0.77& 0.7    &5.17  &5.15 \\
\ \Ha          &  3.       & 3.    &3.         & 2.96&3.  &2.96   & 3.   & 3.    & 3. & 3.     &-     &3.  \\
\ [NII]6585    &  0.7      & 0.76  &0.57       & 0.56&0.85&0.85   & 1.8  & 1.8   &0.78& 0.8    &-     &-\\
\ [SII]6717    &  0.66     & 0.7   &1.19       & 1.1 &0.63&0.66   & 0.27 & 0.5   &-   & -      &-     &-\\
\ [SII]6731    &  0.33     & 0.6   &0.56       & 0.97&0.5 &0.59   &  0.5 & 0.45  &-   & -      &-     &-\\
\   \Vs (\kms) &   -       & 140   &   -       &150  &  - &150    & -    & 140   &-   & 120    &-     &150\\
\  \n0  (\cm3) &  -        & 120   &   -       &130  &  - &130    & -    & 120   &-   & 100    &-     & 100\\
\  \B0  (10$^{-4}$G)&  -   & 3     &   -       &3    &  - &3      & -    & 3     &-   & 2      &-     & 1\\
\ $D$  $^4$    &  -        & 5.3   &   -       &1    &  - & 1     & -    & 1     &-   & 3      &-     & 0.2\\
\ O/H $^5$     &  -        & 6.6   &   -       &6.3  &  - & 6.3   & -    & 6.6   &-   & 6.5    &-     & 6.6\\
\ N/H $^5$     &  -        & 0.2   &   -       &0.2  &  - &0.3    & -    & 0.5   &-   & 0.3    &-     &0.13 \\
\ S/H $^5$     &  -        & 0.04  &   -       &0.1  &  - &0.06   & -    & 0.03  &-   & 0.03   &-     &0.03\\
\ \Ts (10$^4$K)&  -        &3.6    &   -       &3.5  &  - &3.5    & -    & 3.6   &-   &3.8     &-     & 8.3\\
\ $U$          & -         &0.014  &   -       &0.01 & -  &0.01   & -    & 0.014 & -  &0.007   &-     &0.007\\
\ \Hb $^6$     &  -        & 0.014 &   -       &0.016&  - &0.016  & -   &  0.03  & -  & 0.014  &-     & 0.007\\ \hline
\end{tabular}}

$^1$de Ugarte Postigo et al (2014); $^2$ Cucchiara et al (2013); $^3$ Soderberg et al (2006);
$^4$ in 10$^{18}$ cm; $^5$ in 10$^{-4}$; $^6$ in \erg

\end{table*}

\section{Results}

 Most of the  present work  refers to the modelling of the SN and GRB host galaxy line spectra.
The results show that to reproduce narrow line spectra the calculations should account for the
coupled effect of shocks and  radiation  from the starburst within the host galaxy. 
The precision of modelling is given in graphical form for the strongest lines in Fig. 4.
For  [NII]/\Hb the fit of calculated to observed  line ratios is less precise. 
In fact, at the velocities corresponding
to the FWHM observed in the line profiles, the [NII] doublet is blended with the \Ha line.
Therefore, the derived line intensities  are  more uncertain.

Most of the  parameters which  partly   describe
high redshift issues, such as SFR, stellar masses, element abundances, etc.   are calculated from the 
data observed at Earth. On the other hand,    modelling results  reveal the conditions of gas and dust
at the emitting  clouds. 

\subsection{Physical parameters}

The results of shock velocities, preshock densities, cloud geometrical thickness, SB effective 
temperatures, ionization parameters and \Ha absolute fluxes
calculated at the nebula are shown  as function of z in Figs. 5 and 6.
Fig. 5 top diagram shows  the pre-shock densities and the shock velocities calculated in the host clouds.
The ranges for SLSN (\n0 between 100 and 1000 \cm3, and \Vs between 100 and 1000 \kms) are similar to those found in
local AGN and SB galaxies. 
The results  calculated by RD  models are  adapted to  the expanding shocks resulting from the explosion,  
propagating through  the ISM clouds.
\Vs upper limits  calculated by SD models for the Leloudas et al sample of SLSNII hosts
are suitable to the reverse shock. 
The \Hb  fluxes (Table 2)  calculated by SD models  are relatively low and indicate that they could   give only a
minimum contribution to the spectra. Velocities of 1000 \kms are characteristic of the Crab nebula
SNR. Lower \Vs are more appropriated to the older Cygnus Loop SNR filaments.
 In the redshift range where both  LGRB and SN host spectra are observed
 \Vs are higher for the SN hosts.
In particular for LGRB hosts, \n0 and \Vs are located  at the lower limit of the ensemble of the data.
This  suggests that the LGRB  event  negligibly affects   the host galaxy cloud conditions.

Fig. 5 bottom diagram refers to the SB effective temperature and ionization parameter.  
\Ts  are  relatively high in  few
SLSNR hosts; in more SLSNI hosts \Ts $>$ 10 $^5$ K are even higher than in galaxies  showing activity at z$\geq$ 2.
$U$  has a smoother trend.
 LGRB  hosts correspond to relatively low SB temperatures ($\sim$ 3 10$^4$K),  increasing towards  z=2.
The same trend is shown for $U$,  
 which  appears at  the lower limit  for 0.4$<$z$<$ 1.3, 
it is  largely scattered at z$>$ 2, reaching $U \sim$1.
 Such high $U$  reveal a small dilution
 factor of the SB flux.

In the  bottom diagram  of Fig. 6 the cloud geometrical thickness appears as a function of z. 
Except for a few SLSN where $D$  are close to
3-30 pc, SN and GRB  show $D$ $\sim$ 0.3 pc for all the host clouds. At z between 0.3 and 2.3, $D$ splits between
 0.3 pc for GRB and $D<$ 0.003 pc for the other galaxy types. 
$D$  is obtained phenomenologically by the calculations.  It is a leading parameter in modelling because
 linked with flux absorption and emission  in each slab of the downstream gas
at different wavelengths.
Moreover, $D$ determines the emission measure EM=${\Sigma}$ n \ne ${\Delta}$x, where ${\Delta}$x is the
 gas slab thickness ($D$=${\Sigma}$${\Delta}$x) and n and \ne the H density and the electron density, respectively
 in the slabs
   where the physical conditions are  approximately constant. EM enters in the calculations
of  the line intensities.
$D$ depends on fragmentation due to the turbulent regime created by shocks.
Blanchard et al (2015) refer to the galaxy sizes, which in compact galaxies  could roughly represent  the upper  limit to
the geometrical thickness of the clouds.  However, it is generally found that $D$ is small compared with  the 
galaxy dimensions and  several clouds   are included in each galaxy.

\subsection{Metallicities}

Following Vergani et al, metallicity is one of the  main parameters  which affects the evolution of massive stars as
well as their explosive deaths  (e.g. Woosley 1993, etc).
In Fig. 6  top diagram we show the results of O/H and N/H calculated in each host by detailed modelling.
O/H is depleted in most of the SN and SLSN host clouds for 0.1$<$z$<$1, while  it is close to solar in most 
of the GRB hosts.
Regarding the N/H ratio, we note a large distribution of N/H from solar to lower than solar    by a 
factor $>$10 for both SLSN and GRB hosts.
Subsolar N/H is  an indication of external gas acquisition through merger processes.

In Fig. 7   the results of 12+(O/H)$_d$ calculated  for nearby SN Ic hosts by Modjaz et al using Kewley \& Dopita (2002)
method,
SLSN hosts by Leloudas et al.  using the Kobulnicky \& Kewley (2004) method, SN Type Ibc by Sanders et al 
using the Zaritsky et al (1994) method, 
are compared with the results calculated by detailed modelling (12+(O/H)$_m$. 
 Fig. 7 shows that  the results for SLSN hosts roughly  correlates with those calculated by detailed modelling.
 The results for the host galaxies of long and short period  GRB, LGRB which 
contain  WR stars and other LGRB are neglected because  metallicities calculated by detailed modelling
 are near solar for all of them.
  Some observed LGRB hosts show high metallicity. Kr\"{u}ler et al (and references therein)  point out that
"several metal rich GRB were discovered". Graham et al (2015) recently confirmed that high metallicities 
can be found in LGRB.
It was explained by Contini (2014a) that the metallicities in terms of the O/H  relative abundances
obtained by "direct methods" are lower limits. 

 Our results  about near solar O/H in most LGRB hosts  are  different from  those generally obtained by other modelling 
methods  which indicate low metallicities in LGRB galaxy hosts. This 
leads to serious consequences  in evolution theories based on metallicities.
We justify our results by claiming  that the  theoretical calculations of 
spectral lines and continuum    were carefully  checked by   modelling the rich spectra of local
objects. On the other hand, "direct methods" and their results are universally accepted by the
scientific community.

\begin{figure}
\centering
\includegraphics[width=8.0cm]{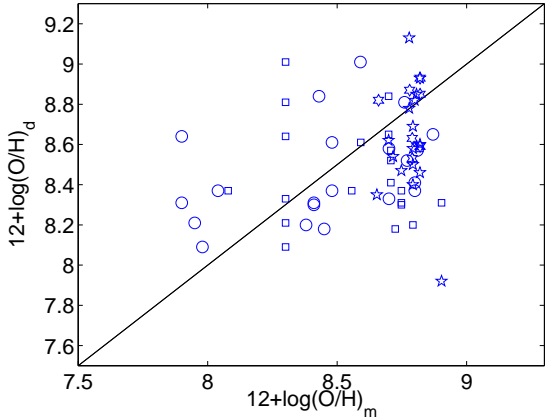}
\caption{Comparison of O/H  calculated by detailed modelling (m) with
those calculated by direct methods (d) by Leloudas et al :  circles (SB models),  squares
(SD models); 
by Modjaz et al :  hexagrams;
by Sanders et al : pentagrams.}
\end{figure}

\subsection{SFR, \Ha line flux and the gas physical parameters}

SFR  are generally calculated by the observers from the \Ha luminosity \La.
 We will limit our discussion on the SFR, comparing the SFR trend  with that of the calculated \Ha fluxes. 
The \Ha flux calculated at the nebula is shown in the   bottom diagram (top panel) of Fig. 6.
The low \Ha fluxes calculated by SD models in SLSN host clouds  are easily lost in the FWHM sockets of  line profiles.
The trend of \Ha for  a few SLSNI at lower z is similar to the trend of \Ts and $U$ calculated by  the RD models, 
indicating that in the high \Ts range the \Ha fluxes depend on \Ts and $U$.
We added in  the diagram the SFR  calculated by the observers for each host galaxy of  SN,  SLSN
(black open diamond) and LGRB (black open circles) surveys shifted by a factor of 1000 for sake of clarity.
The increasing trend of SFR with z is clear because it depends on \La. The  increase of \Ha calculated at the nebula   
with z is   less evident.  
 Fig. 8 diagrams  show \Ha (calculated at the nebula) as function of \Ts, $U$, \Vs, \n0, 
$D$ and O/H. The top left diagrams  indicate that at relatively low SB temperatures
(\Ts$\leq$ 3 10$^4$ K), \Ha fluxes do not increase with \Ts and LGRB hosts even show a decreasing trend
with  SLSNII objects in the top tail. This unexpected trend is  due to the
effect of  the other   parameters. At high \Ts ($\geq$ 10$^5$K) as those found for SLSN host SB, 
 \Ha increases with \Ts.
A nearly linear increasing trend of log(\Ha)  with log($U$) is evident in Fig. 8 top right diagram.
 \Ha  is a strong function of  the ionization parameter $U$, which depends on both the
photoionization source (SB) flux and  the  host cloud density.
\Ts and $U$ different  effects on the calculated line ratios are  discussed in Sect. 2.2.
The increasing trends of \Ha with \Vs and \n0 appear in the middle left and right diagrams, respectively,
neglecting  SD models for SN.
At the relatively low \Vs ($<$ 350 \kms) and \n0 ($<$ 300 \cm3) which are characteristic of the gas in the
host galaxies presented in previous sections,
the effect of the other parameters leads to the large scattering of the results.
  Accordingly, the increasing trend of \Ha with $D$ (bottom left) is strongly disturbed.
The trend of \Ha with  O/H (bottom right) is hardly evident for SLSN hosts which are distributed on
a relatively large O/H range. It is even less clear for GRB hosts because most of them show solar O/H.

SFR are  calculated by the \Ha luminosity observed at Earth \La  (Kennicutt 1998).
SFR $\propto$ \La, and \La=\Ha(at Earth)$\times$ 4 $\pi$ d$^2$=\Ha(at the  nebula)$\times$ 4 $\pi$ R$^2$
(adopting a filling factor \ff=1) where d is the 
distance to Earth and R the radius of the cloud, i.e. the distance of the cloud from the host galaxy SB stars.
Fig. 9   shows the results of R as function of z for \ff=1. 
 The emitting cloud radius in the SLSN hosts are lower than in GRB hosts and in SN hosts at lower z,
indicating more compact galaxies. 
The clouds in SGRB  hosts have larger radius than in LGRB hosts at the same z, however,
 the data are too few to be meaningful.
Larger R,  best fitting the galaxy observed dimensions, could be found for \ff $<$ 0.01.
It seems from Fig. 9  that the increasing trend of SFR with z at high z depends  on R rather than on \Ha which
follows the trend of $U$.
New data at higher z are needed.

\begin{table}
\centering
\caption{The symbols in Figs. 5, 6 and 8}
\tiny{
\begin{tabular}{lcccccccccccc} \hline  \hline
\ blue  circles&  SLSNR hosts (Leloudas et al)\\
\ blue   open triangles at z$\geq$0.1 & SLSNII hosts  (Leloudas et al)\\
\ blue  filled triangles & SLSNI hosts  (Leloudas et al)\\
\ blue squares & Shock  models for SLSN hosts (Leloudas et al)\\
\ blue  x     & Type Ic host central spectra (Modjaz et al) \\          
\ blue encircled x & Type Ic hosts  at SN positions (Modjaz et al)\\
\ blue  triangles  at z$\leq$ 0.1 & SN Ib host (Sanders et al)\\
\ blue  encircled triangles  & SN IIb hosts (Sanders et al)\\
\ blue opposite triangles    & SN Ic hosts (Sanders et al)\\
\ blue double triangles      & SN IcBL hosts (Sanders et al)\\
\ blue hexagrams              & SN Ibc (Sanders et al) \\
\ red asterisks &GRB hosts  (Kr\"{u}hler et al)\\
\ red x       & LGRB hosts (Savaglio et al) \\
\ yellow hexagrams & SGRB hosts (de Ugarte Postigo et al)\\
\ green pentagrams  & LGRB different hosts (Table 8)\\
\ magenta plus & LGRB hosts with WR stars (Han et al)\\
\ black dots & other high z galaxies (Contini 2015)\\ \hline.

\end{tabular}
}
\end{table}

\begin{figure*}
\centering
\includegraphics[width=8.6cm]{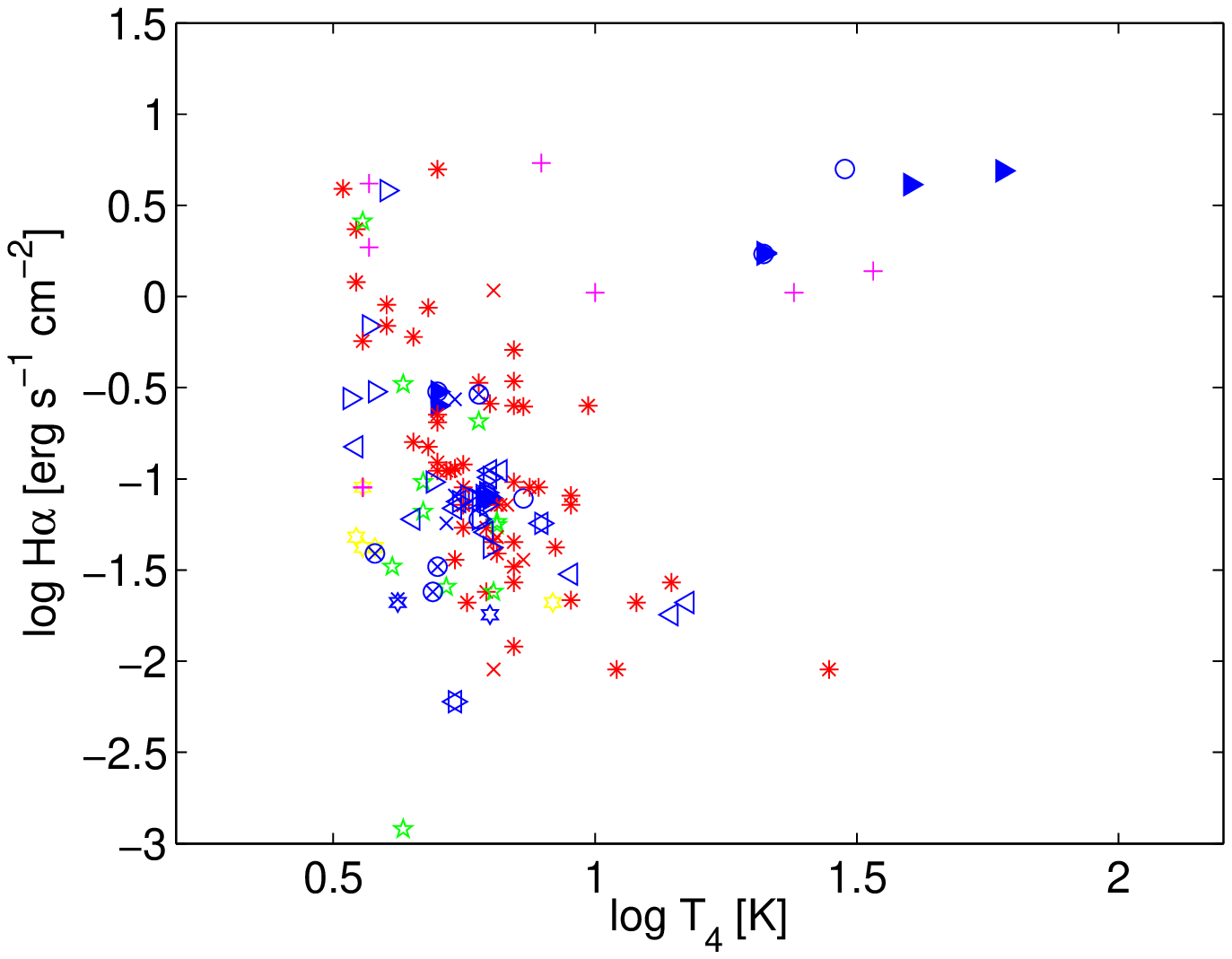}
\includegraphics[width=8.6cm]{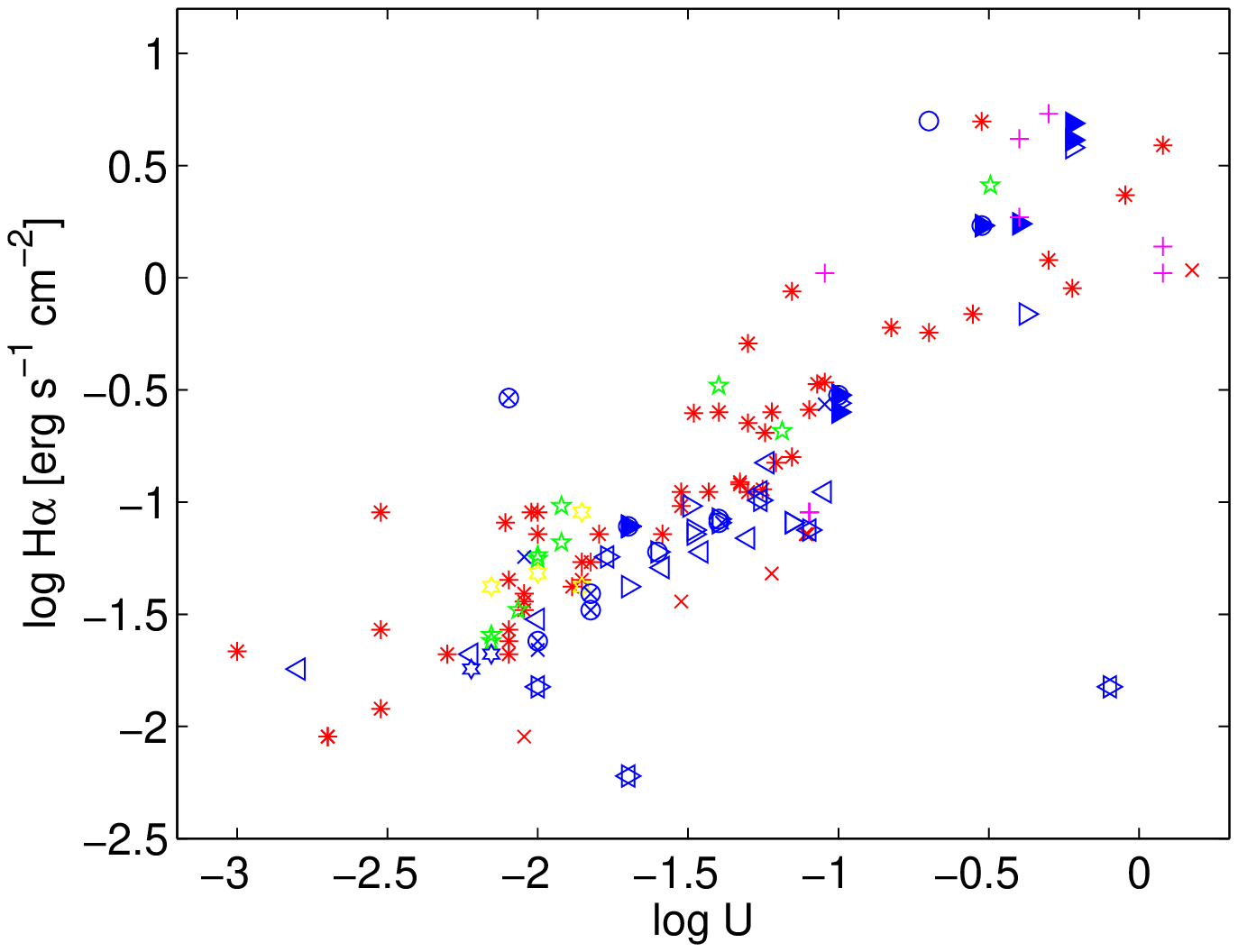}
\includegraphics[width=8.6cm]{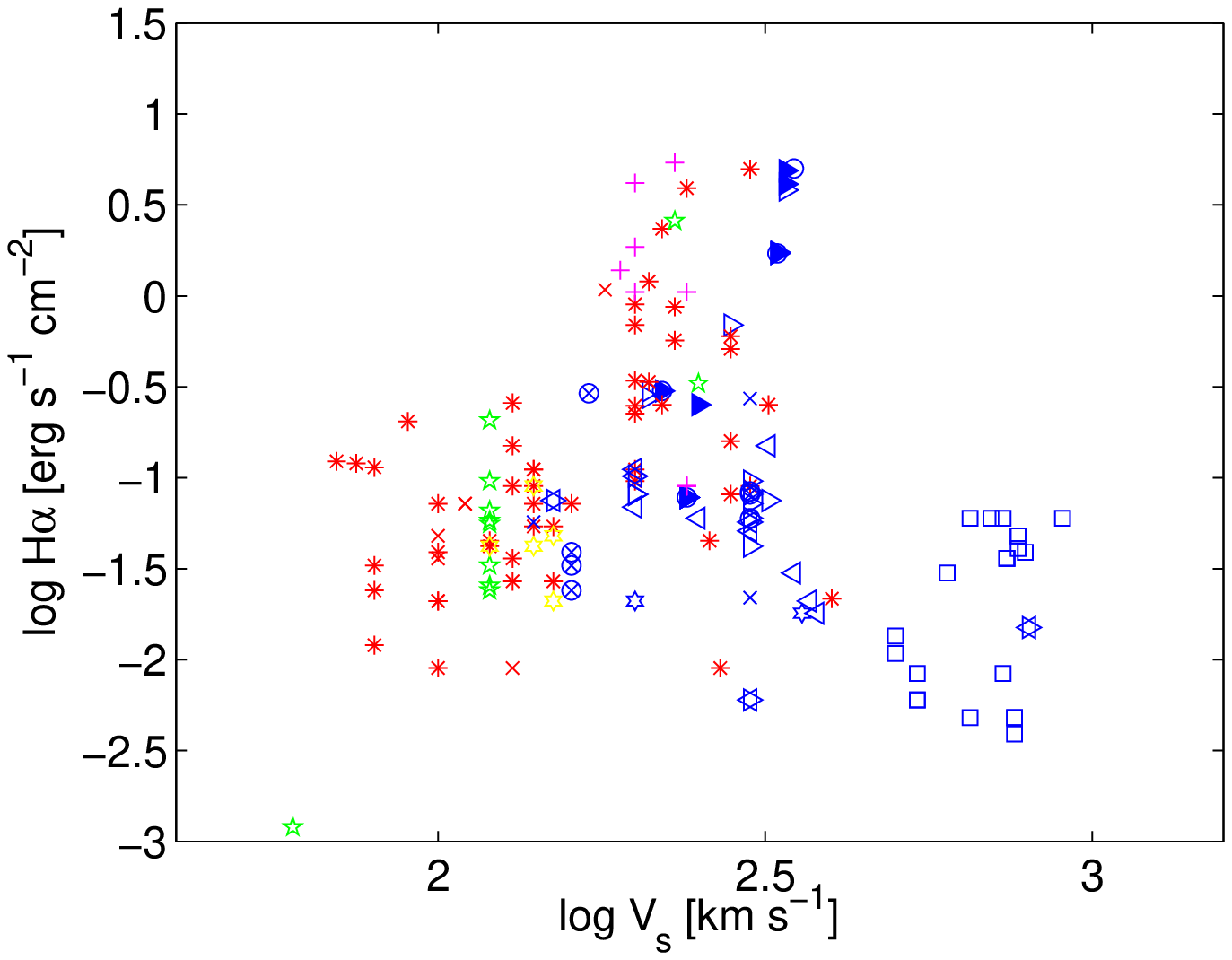}
\includegraphics[width=8.6cm]{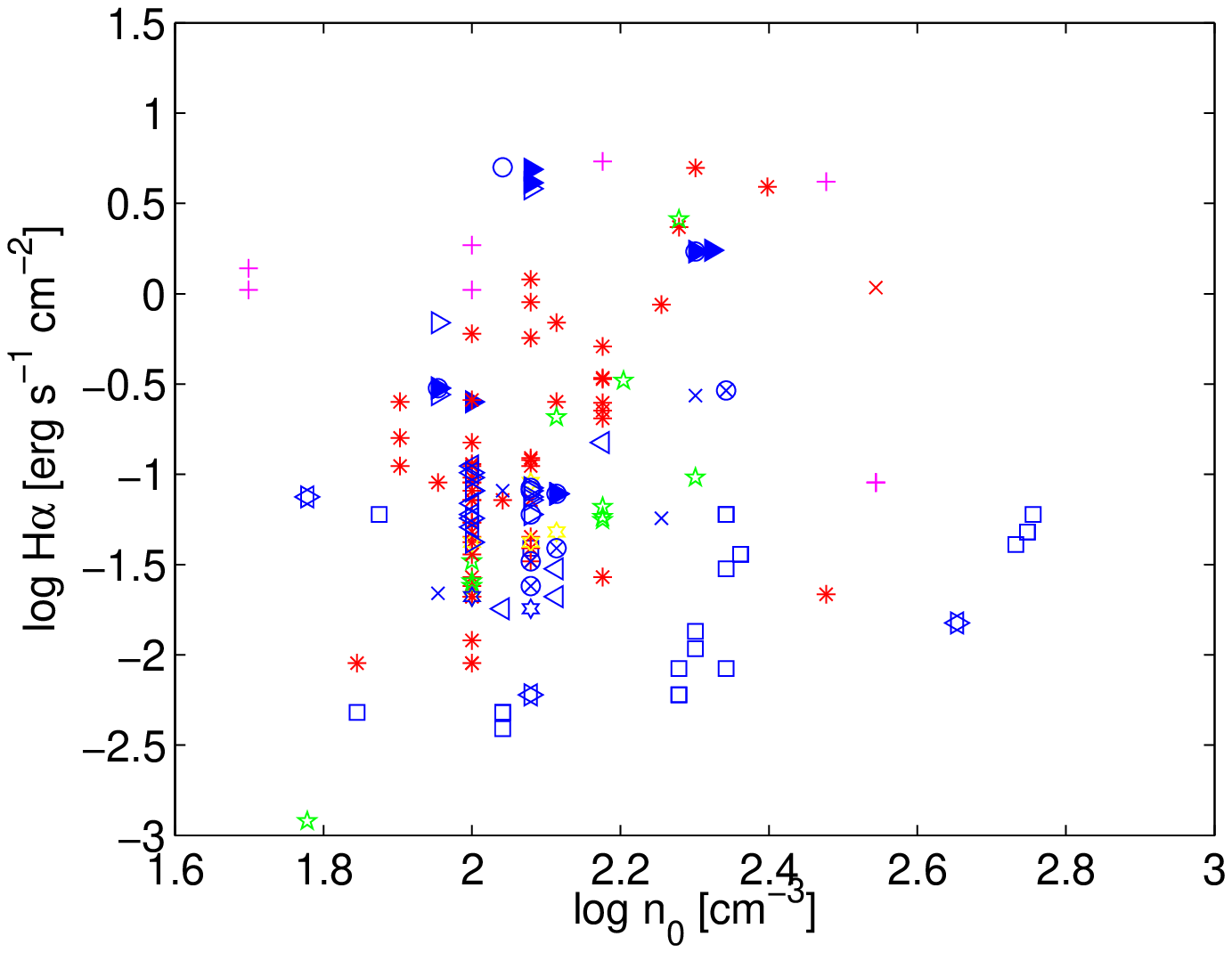}
\includegraphics[width=8.6cm]{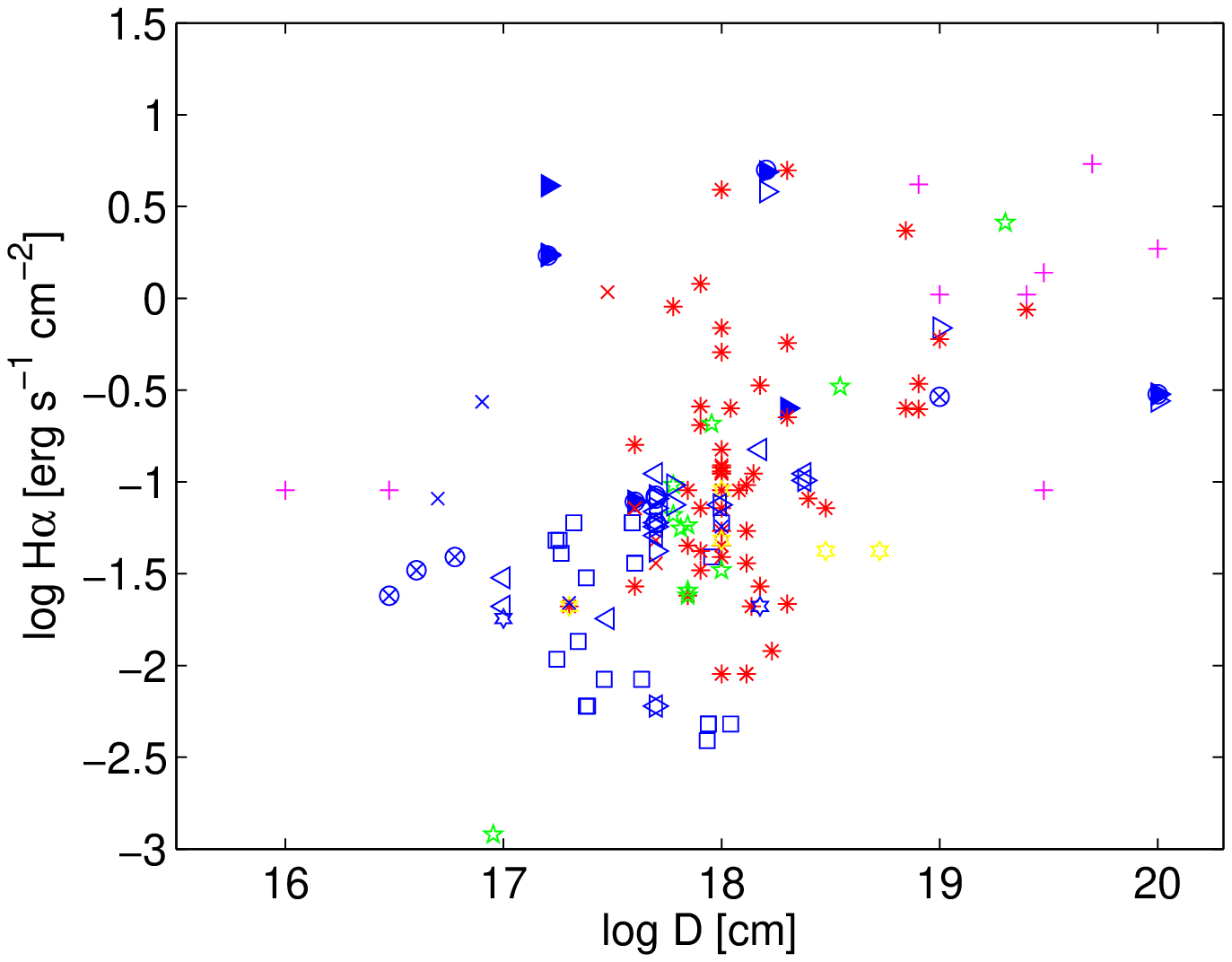}
\includegraphics[width=8.6cm]{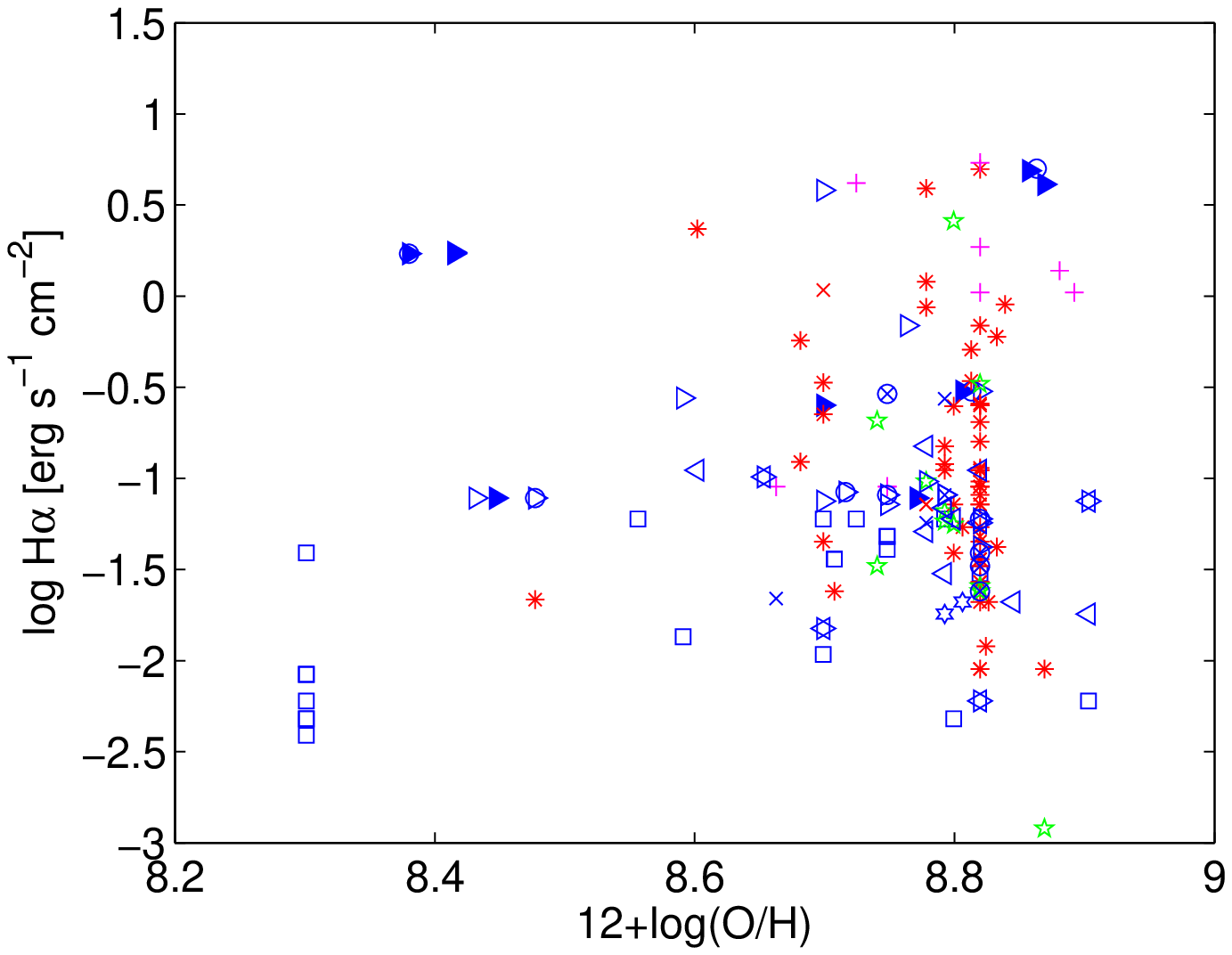}
\caption{\Ha as function of the physical parameters.
Symbols as in Fig. 5}
\end{figure*}

\section{Concluding remarks}

The results  obtained by modelling broad lined SN Type Ic hosts by Modjaz et al, SN hosts by
Sanders et al and   SLSN host galaxies  by Leloudas et al    are
presented in this paper and  compared 
with those  calculated for various  LGRB  and a few SGRB host galaxies  at  z$<$3.4.
It was found that:

1) The differences between the II , I, and R SLSN types are irrelevant in terms of the host \Vs and \n0. 
 but \Ts and $U$ calculated for a few SLSNI hosts  ($>$ 10$^5$K and $\geq $0.1, respectively) 
are higher than for the other
SLSN types. The latter are similar to \Ts and $U$ calculated for 
 LGRB hosts ($\sim$ 3-7 10$^4$K and 0.01-0.1, respectively).

2) The shock velocities and pre-shock densities  in  SN  host galaxies are  similar to those
found in the other  galaxies reported in Fig. 5, at the same z, but  in LGRB hosts they are lower.
The \Vs calculated by shock dominated models  for SLSN hosts are higher than for  radiation dominated 
models, but still lower than those calculated for the  \Ly line emitting galaxies at higher z.
In the GRB host spectra there is no trace of the high velocity winds predicted by SD models
in  SLSN host spectra.
 Such winds would waive the   low density extended clouds in the ISM of these  hosts.

3)  \Ts and $U$ calculated for  LGRB hosts  show that these are rather quiescent galaxies even at z$\sim$ 2.
It was found by Contini (2014b) that activity is associated to a high \Ts.

4) In SN   host  galaxies some SB stars  have temperatures  approaching  those of outburst.

5)  Models reproducing the line spectra of a sample of LGRB galaxies hosting WR stars (Han et al 2010) 
at relatively low redshift,
show  He/H=0.13   in  a few objects. 
Modelling the continuum SED, the contribution of
a rather old stellar population (3-8 10$^3$ K)  appears in the  near-IR - optical  domain.

6) The O/H relative abundances calculated by detailed modelling of SN hosts are scattered between 
12+log(O/H) =8.85 and 8.0. They are close to solar  in LGRB hosts,
 perhaps contaminated by   shocked an photoionised gas  within the same galaxy. 
O/H   are generally higher than those calculated by   "direct methods", 
in particular for LGRB.
N/H  are  lower than solar  by  factors of 5-20 for both SN and GRB hosts.

7) The geometrical thickness of the clouds is nearly constant for GRB hosts,
$D$$\sim$ 0.3 pc  and shows a small increase at z$\sim$ 2. 
A  uniform $D$   indicates similar turbulent regimes in the ISM of the different hosts.
In fact similar shock velocities are  found in  GRB host clouds.

8) The \Ha absolute fluxes calculated at the emitting clouds of SLSNI hosts at 0.1$<$ z$<$0.3,
  are   relatively high,  proportionally to  \Ts and $U$.

9)  SLSN host galaxies are more compact than other SN hosts.

Concluding, our analysis of SLSN and GRB host galaxy spectra indicates 
 that some  parameters adopted to reproduce the SN and LGRB  host galaxy
line ratios are different.
First, we have found that the velocity field is lower in LGRB host than in SN host galaxies, and it is
very similar to that found  in SB galaxies. Second,  most stars in the
SB throughout the SN hosts   have reached temperatures  similar  to  those of outburst, 
indicating that some activity is going on.
Moreover, our modelling leads to lower than solar metallicities in term of  O/H in SN hosts
at 0.1$<$z$<$1. Metallicities  in LGRB hosts are close to solar throughout the 0.01$<$z$<$3.4 range.
The present analysis results of SN and GRB host galaxy spectra, in the near UV-optical-far IR range, suggest 
that the SN-host symbiosis is stronger than the GRB-host one in terms of activity.
The physical and chemical conditions in the GRB host galaxies are similar to those in SB galaxies within a large
z range.

\begin{figure}
\centering
\includegraphics[width=9.2cm]{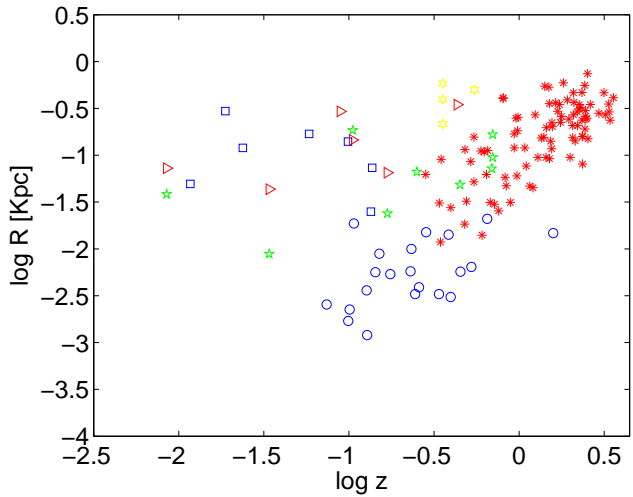}
\caption{The calculated radius of the emitting clouds  
as a function of  redshift. 
Circles : SLSN (Leloudas et al);  squares : SN Type Ic (Modjaz et al);
  asterisks : LGRB (Kr\"{u}hler et al);  triangles :LGRB  (Savaglio et al);
 stars  refer to LGRB from  surveys  including Han et al data;  
hexagrams : SGRB.
}

\end{figure}

\section*{Acknowledgements}
I am very grateful to the referee for critical suggestions which substantially
improved the presentation of the paper.

\section*{References}

\def\ref{\par\noindent\hangindent 18pt}

\ref Allen, C.W. 1976 Astrophysical Quantities, London: Athlone (3rd edition)
\ref Anders, E., Grevesse, N. 1989, Geochimica et Cosmochimica Acta, 53, 197
\ref Anderson, J.P. et al 2016 arXiv160200011A
\ref Asplund, M., Grevesse, N., Sauval, A.J., Scott, P. 2009, ARAA, 47, 481
\ref Berger, E.  et al 2005, Nature, 438, 988  
\ref Blanchard, P.K. et al 2015 arXiv:1509.07866
\ref Castro-Tirado, A.J. et al 2001, A\&A,  370, 398 
\ref Chevalier, R.A. 1982, ApJ, 259, L85
\ref Christensen, L., Vreeswijk, P.M., Sollerman, J. et al 2008, A\&A, 490, 45
\ref Contini, M. 2016, submitted
\ref Contini, M. 2015, MNRAS, 452, 3795
\ref Contini, M. 2014b, A\&A, 572, 65
\ref Contini, M. 2014a, A\&A, 564, 19
\ref Contini, M. 2009, MNRAS, 399, 1175
\ref Contini, M. 2004, A\&A, 422, 591
\ref Contini, M. 1987, A\&A, 183, 53
\ref Contini, M. , Shaviv, G. 1980 ApSS, 85, 203
\ref Contini, M., Kozlovsky, B.Z., Shaviv, G. 1977, A\&A, 59, 387
\ref Cox, D.P. 1972, ApJ, 178, 143
\ref Cucchiara, A. et al. 2013, ApJ, 777,94
\ref de Ugarte Postigo, A. et al 2014, A\&A, 563, 62
\ref Fesen, R., Milisavljevic, D. 2015, AAS, 22514024 
\ref Filippenko, A.  1997, ARA\&A, 35, 309
\ref Fishman, G.J., Meegan, C.A. 1995 ARA\&A, 33, 415
\ref Fong, W. et al 2013, ApJ, 769, 56
\ref Fruchter, A.S. et al. 2006, Nature, 441, 463
\ref Gal-Yam, A. et al 2004, ApJ, 609, L59
\ref Graham, J.F. et al 2015 arXiv:1511.00667v
\ref Graham, J.F., Fruchter, A.S. 2013, ApJ, 774, 119
\ref Hammer, F., Flores, H., Schaerer, D., Dessauges-Zavadsky, M., Le Floc'h, E., Puech, M. 2006, A\&A, 454, 103	
\ref Han, X.H., Hammer, F., Liang, Y.C., Flores, H., Rodrigues, H., Hou, J.L., Wei, J.Y. 2010, A\&A, 514, A24
\ref Hjorth, J. et al. 2003, Nature, 423, 847
\ref Iwamoto, K. et al. 1998, Nature, 395, 672
\ref Kann, D.A. et al ApJ, 734, 96
\ref Kennicutt, Jr. R.C. 1998, ARA\&A, 36, 189
\ref Kewley, L.J., Dopita, M.A. 2002 ApJS, 142, 35
\ref Kobulnicky, H.A., Kewley, L.J. 2004, ApJ, 617, 240
\ref Kouveliotou, C. et al 1993, ApJ, 413, L101
\ref Kriek, M. et al 2009, ApJ, 700, 221
\ref Kr\"{u}hler, J. et al 2015 A\&A, arXiv:1505.06743
\ref Leloudas, G. et al 2015, MNRAS, 574, A61
\ref Levesque, E.M., Berger, E., Kewley, L.J., Bagley, M.M. 2010, ApJ, 139, 694
\ref Maoz, D., Mannucci, F., Nelemans, G. 2014, ARA\&A, 52, 107
\ref Modjaz, M. et al 2008, AJ, 135, 1136
\ref Nagao, T., Maiolino, R. , \& Marconi, A. 2006, A\&A, 459, 85
\ref Osterbrock, D.E. 1974 in 'Astrophysics of Gaseous Nebulae' W.H. Freeman and Company, San Francisco
\ref Paczynski, B. 1998 AIPC, 428, 783
\ref Piranomonte, S. et al 2015, MNRAS, 452, 3293
\ref Rodriguez-Ardila, A., Contini, M., Viegas, S.M. 2005, MNRAS, 357, 220
\ref Rosswog, S.,Ramirez-Ruiz, E., Davies, M.B. 2003, MNRAS, 345,1077
\ref Sanders, N.E. et al. 2012 ApJ, 758, 132
\ref Savaglio, S., Glazerbrook, K., Le Borgne, D. 2009, ApJ, 691, 182
\ref Soderberg, A.M.  et al 2006, ApJ, 650, 261
\ref Sollerman J.; \"{O}stlin, G.; Fynbo, J. P. U.; Hjorth, J.; Fruchter, A.; Pedersen, K.
	2005, NewA, 11, 103
\ref Stanek, K.Z. et al 2003, ApJ, 591, L17
\ref Van Dokkum, P.G., Kriek, M., Rodgers, B., Franx, M., Puxley, P. 2005, ApJ, 622, L16
\ref Vergani, S.D. et al 2011 A\&A 535, A127
\ref Williams, R.E. 1967, ApJ, 147, 556
\ref Woosley, S.E. 1993, ApJ, 405, 273
\ref Zaritsky, D., Kennicutt, R.C. Jr, Huchra, J.P. 1994, ApJ, 420, 87
\end{document}